\documentclass[twocolumn,nofootinbib,notitlepage,showpacs,prx,amsmath,amstex,amssymb,citeautoscript,longbibliography]{revtex4-2}
\pdfoutput=1
\usepackage{natbib}
\usepackage[english]{babel}
\usepackage{letltxmacro}
\usepackage{latexsym}
\usepackage[dvipsnames]{xcolor}
\LetLtxMacro{\ORIGselectlanguage}{\selectlanguage}
\makeatletter
\DeclareRobustCommand{\selectlanguage}[1]{%
  \@ifundefined{alias@\string#1}
    {\ORIGselectlanguage{#1}}
    {\begingroup\edef\x{\endgroup
       \noexpand\ORIGselectlanguage{\@nameuse{alias@#1}}}\x}%
}
\newcommand{\definelanguagealias}[2]{%
  \@namedef{alias@#1}{#2}%
}
\makeatother
\definelanguagealias{en}{english}
\definelanguagealias{English}{english}
\usepackage{graphicx}
\usepackage{amsmath}
\usepackage{amsfonts}
\usepackage{amssymb}
\usepackage{bm}
\usepackage{color}
\usepackage{soul} 
\usepackage{amssymb}
\usepackage{wasysym}
\usepackage{comment}
\usepackage{dsfont}
\usepackage{float}
\usepackage{braket}
\usepackage{hyperref}
\hypersetup{
    bookmarks=false,         
    unicode=false,          
    pdftoolbar=false,        
    pdfmenubar=true,        
    pdffitwindow=false,     
    pdfstartview={FitH},    
    pdftitle={Discrete time crystalline order enabled by quantum many-body scars: entanglement steering via periodic driving},    
    pdfauthor={N. Maskara et al.},     
    pdfsubject={quantum scars},   
    pdfcreator={},   
    pdfproducer={}, 
    pdfkeywords={}, 
    pdfnewwindow=true,      
    colorlinks=true,       
    linkcolor=black,          
    citecolor=blue,        
    filecolor=magenta,      
    urlcolor=blue           
}
\newcommand{\be}{\begin{equation}}
\newcommand{\ee}{\end{equation}}
\newcommand{\bea}{\begin{eqnarray}}
\newcommand{\eea}{\end{eqnarray}}

\newcommand{\tr}{\mathop{\rm tr}}

\newcommand{\NAB}{{\cal I}}
\newcommand{\titleofthepaper}{Discrete time-crystalline order enabled by quantum many-body scars: entanglement steering via periodic driving}
\begin{document}
\title{\titleofthepaper}
\date{\today}
\author{N. Maskara$^{1}$}
\author{A.~A.~Michailidis$^{2}$}
\author{W. W. Ho$^{1,3}$}
\author{D. Bluvstein$^{1}$}
\author{S. Choi$^{4,5}$}
\author{M. D. Lukin$^{1}$}
\author{M. Serbyn$^{2}$}
\affiliation{$^1$Department of Physics, Harvard University, Cambridge, MA 02138, USA}
\affiliation{$^2$IST Austria, Am Campus 1, 3400 Klosterneuburg, Austria}
\affiliation{$^3$Department of Physics, Stanford University, Stanford, CA 94305, USA}
\affiliation{$^{4}$Department of Physics, University of California Berkeley, Berkeley, CA 94720, USA}
\affiliation{$^{5}$Center for Theoretical Physics, Massachusetts Institute of Technology, Cambridge, MA 02139, USA}
\begin{abstract}
The control of many-body quantum dynamics in complex systems is a key challenge in the quest to reliably produce and manipulate large-scale quantum  entangled states. Recently, quench experiments in Rydberg atom arrays (Bluvstein \emph{et al.} \href{https://science.sciencemag.org/content/early/2021/02/24/science.abg2530}{\emph{Science},  25 Feb 2021}) demonstrated that coherent revivals associated with quantum many-body scars can be stabilized by periodic driving, generating stable subharmonic responses over a wide parameter regime. We analyze a simple, related model where these phenomena originate from spatiotemporal ordering in an effective Floquet unitary, corresponding to discrete time-crystalline (DTC) behavior in a prethermal regime. Unlike conventional DTC, the subharmonic response exists only for N\'eel-like initial states, associated with quantum scars. We predict robustness to perturbations and identify emergent timescales that could be observed in future experiments. Our results suggest a route to controlling  entanglement in interacting quantum systems by combining periodic driving with many-body scars.  
\end{abstract}
\maketitle

{\it Introduction.}---Creating and manipulating entanglement is a fundamental goal of quantum information science, with broad implications in computation, metrology, and beyond. At the same time, not all forms of entanglement are useful. In particular, strongly interacting quantum many-body systems  generate large amounts of entanglement under their intrinsic dynamics, in a process known as thermalization~\cite{Polkovnikov-rev,Kaufman2016}. However, such dynamics irreversibly scramble local quantum information, erasing memory of the initial state. Creating and controlling entanglement while at the same time combating thermalization~\cite{AbaninRMP,Huse-rev,SerbynQMBS20} in isolated interacting many-body systems~\cite{Bloch2008,quantumsimulationRMP14,Browaeys2020} is therefore essential for applications of large-scale entangled states~\cite{Horodecki,Pezze}. 

Experimental studies involving programmable quantum simulators based on Rydberg atom arrays~\cite{Bernien2017} have suggested that interacting quantum systems can exhibit a weak breakdown of thermalization, where certain initial conditions exhibit surprising, persistent many-body revivals. This phenomenon can be viewed as resulting from so-called quantum many body scars (QMBS)~\cite{Turner2017,wenwei18TDVPscar}-- anomalous, non-thermal many-body eigenstates -- named in analogy to non-ergodic wavefunctions in the spectrum of otherwise chaotic single particle Hamiltonians~\cite{Heller84}. Intriguingly, in some models with QMBS the system undergoes periodic entanglement and disentanglement cycles~\cite{wenwei18TDVPscar,Choi2018,Michailidis2020,Chattopadhyay}, providing a potential route to the controlled manipulation of entanglement dynamics. In practice, however, QMBS are  fragile~\cite{Turner2017,Choi2018,Khemani2018,Lin2020}; since they rely on a dynamically disconnected subspace of non-thermalizing eigenstates~\cite{Choi2018,ShiraishiMori,SerbynQMBS20}, additional interactions generically lead to thermalization~\cite{Lin2020}.

Recent experiments~\cite{Dolev20} demonstrated that periodic driving can dramatically increase the lifetime of scarred oscillations. This observation is surprising, since the experiments used driving frequencies resonant with the local energy scales of the system, permitting easy energy absorption and rapid heating towards a featureless, infinite-temperature state. Additionally, the experiment observed a robust subharmonic response at half of the driving frequency, suggestive of discrete time-crystalline (DTC) order~\cite{Khemani16,Else16}.

In this Letter, we propose a theoretical framework for understanding these experimental observations by introducing a mechanism whereby driving stabilizes quantum scarred oscillations, prolonging their lifetime and protecting them against arbitrary perturbations. Specifically, we focus on the simplest model describing Rydberg blockade -- the so-called PXP model~\cite{FendleySachdev,Lesanovsky2012,Bernien2017} -- which is an idealized model for the Rydberg atom array experiment, with the addition of kicked driving~\cite{Dolev20}. This model exhibits robust subharmonic responses and many-body revivals coming from an effective many-body spin echo. The deviation from a perfect echo introduces a small parameter, allowing us to derive an effective prethermal description of the Floquet dynamics, and argue for stability up until parametrically long times~\cite{Else2017_Prethermal}. 

Namely, we construct an effective Hamiltonian in a rotating frame, hosting an emergent $\mathds{Z}_2$ symmetry which is spontaneously broken in its gapped ground state manifold. When viewed in the laboratory frame, the system oscillates between the two spontaneously broken ground states, resulting in a robust subharmonic response characteristic of DTC~\cite{vonKeyserlingk2016,Else2017_Prethermal,Yao_2017}. However, this subharmonic response is restricted only to N\'eel-like initial states which have a strong overlap with the ground state of the effective Hamiltonian --- a property inherited from QMBS.  Our model differs crucially from earlier works on homogenous time crystals in 1D~\cite{Huang_2018_CFTC,Pizzi20,Mukherjee2020,Yarloo20} and mean-field constructions~\cite{Russomanno_2017}, in that the trajectory being stabilized is generated by an interacting Hamiltonian, which produces non-trivial entanglement. Therefore, our construction opens a prospective route towards coherent control of entanglement dynamics.

{\it Model and phenomenology.}---We study a periodically kicked model 
$H(t)$\,$=$\,$H_\text{PXP}$\,$+$\,$\theta N \sum_{k \in \mathds{Z}} \delta({t-k\tau})$, which generates the following one-period Floquet unitary,
\begin{eqnarray}\label{Eq:UF}
    U_F(\theta ,\tau) &=& e^{-i\theta N} e^{-i \tau H_\text{PXP}},\\ \label{Eq:HPXP}
        H_\text{PXP}&= & \sum_{i=1}^L P_{i-1}\sigma^x_i  P_{i+1}, \quad N = \sum_{i=1}^L n_i,
\end{eqnarray}
describing evolution with the PXP Hamiltonian $H_\text{PXP}$~\cite{FendleySachdev,Lesanovsky2012,Bernien2017}  for time $\tau$, followed by   the number operator $N$ applied through rotation angle $\theta$. For simplicity, the model is defined on a 1D chain of $L$ sites with periodic boundaries, although much of the analysis carries over to higher dimensional bipartite lattices. Each site is a two-level system spanned by a ground (${\circ}$) and an excited (${\bullet}$) state, and periodic boundary conditions are assumed unless stated otherwise. Operators $n_i = \ket{{\bullet}}\bra{\bullet}_i$ and $P_i=\ket{{\circ}}\bra{\circ}_i$ project a given site onto the excited and ground states respectively, while $\sigma^x_i = \ket{{\circ}}\bra{\bullet}_i+\ket{{\bullet}}\bra{\circ}_i$ generates Rabi oscillations. In the Hamiltonian, $\sigma^x_i$ is dressed by projectors on neighboring sites, ensuring that dynamics remain within the blockaded subspace where adjacent sites are never simultaneously excited.

\begin{figure}[t]
\begin{center}
\includegraphics[width=0.99\columnwidth]{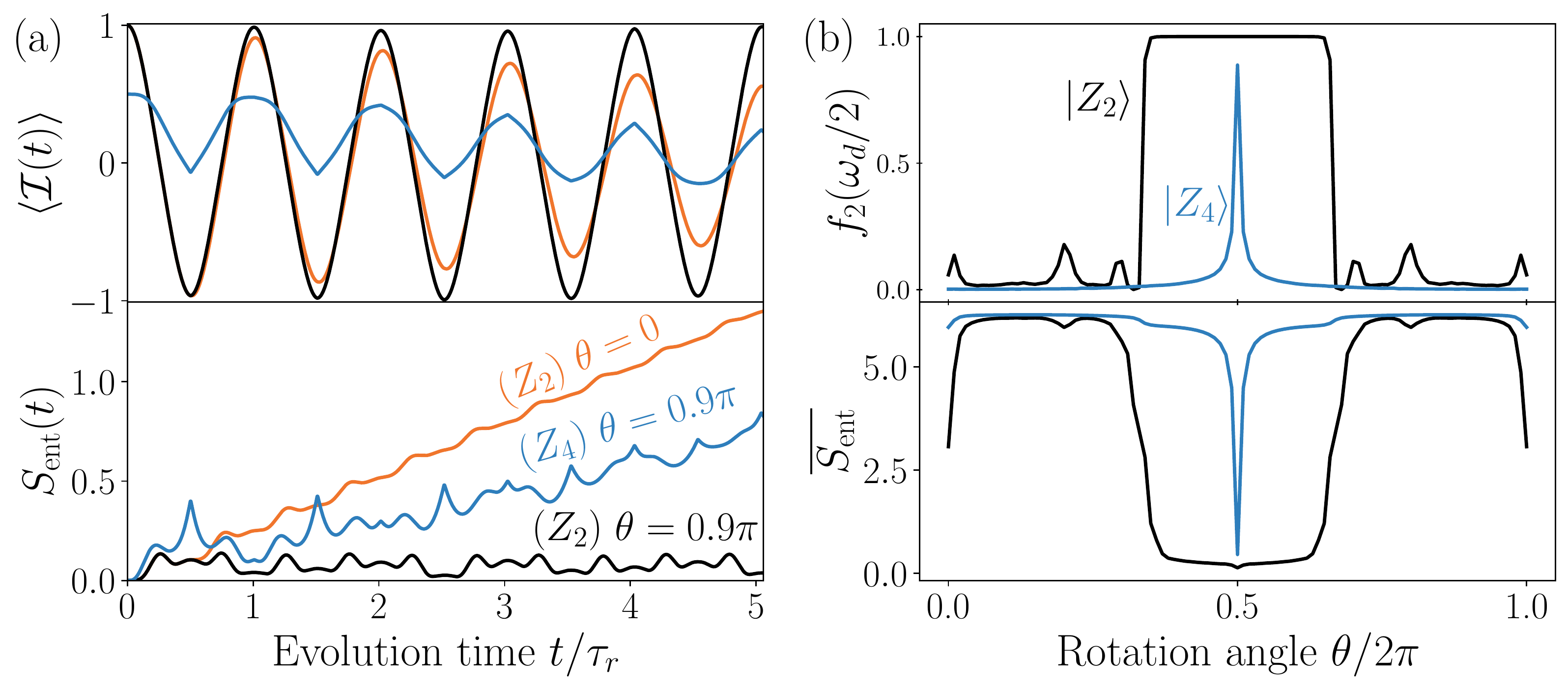}
\caption{ \label{Fig:pheno}
(a) The density imbalance $\NAB$ and bipartite entanglement entropy characterize oscillations between the two N\'eel ordered states ($\ket{Z_2}$), colored orange (undriven) and black (driven), in an infinite size chain simulated via iTEBD~\cite{SOM}. Adding driving with $\tau = 0.993 \tau_r/2$ and $\theta = 0.9\pi$ to PXP model arrests the growth of entanglement entropy $S_{\rm ent}$ and prolongs the lifetime. In contrast, driven dynamics from the $\ket{Z_4} = \ket{{\bullet}{\circ}{\circ}{\circ}{\bullet}{\circ}{\circ}{\circ}\ldots}$ state thermalize rapidly (blue). 
(b)~Subharmonic weight and average entanglement entropy, computed over 400 cycles ($T=400\tau$), for an $L=28$ chain, over a range of $\theta$ and $\tau=0.993 \tau_r/2$. From the $\ket{Z_2}$ state, these observables form a stable plateau around $\theta = \pi$. However from the $\ket{Z_4}$ state, which we use as a stand-in for a generic state, the response quickly disappears away from the effective echo point at $\theta = \pi$.
}
\end{center}
\end{figure}

For $\theta=0$ the Floquet dynamics~(\ref{Eq:UF}) is equivalent to undriven evolution under $H_{\text{PXP}}$. The PXP model is non-integrable and features rapid growth of bipartite entanglement entropy, $S_\text{ent}(t) = -\tr \rho\ln \rho$ where $\rho$ is the half-chain density matrix, from the majority of product states. In contrast, quenching from the N\'eel state $\ket{Z_2} = \ket{{\bullet}{\circ}{\bullet}{\circ}\ldots}$ leads to coherent oscillations between $\ket{Z_2}$ and its inversion partner $\ket{Z_2'}$, as first seen in~\cite{Bernien2017}, with oscillation period $\tau_r \approx 1.51\pi$ that sets an intrinsic resonant timescale. These oscillations are captured by the imbalance in excitation number between odd and even sites, $\NAB =(2/L)\sum_{i=1}^{L/2} (n_{2i-1}-n_{2i})$, see Fig.~\ref{Fig:pheno}(a). However,  dynamics under $H_\text{PXP}$ still generate entanglement, and the coherent many-body oscillations eventually decay, see Fig.~\ref{Fig:pheno}(a).

The addition of strong driving with $\theta \approx \pi$ almost completely suppresses thermalization at early times, most clearly seen in the nominal growth of entanglement entropy over multiple cycles, see Fig.~\ref{Fig:pheno}(a). Concomitantly, oscillations of $\NAB$ synchronize to half the drive frequency, a phenomenon known as subharmonic locking.  The origin of this response is related to the existence of a special point at $\theta = \pi$,  because $H_{\text{PXP}}$ anticommutes with the operator ${\cal C}=\prod_i \sigma^z_i = e^{-i \pi N}$, corresponding to a ``particle-hole symmetry''~\cite{Turner2017}. 
Indeed, driving with $\theta=\pi$ implements an effective many-body echo, since $U_F(\pi,\tau)^2 = {\cal C} e^{-i\tau H_\text{PXP}}{\cal C} e^{-i\tau H_\text{PXP}} = e^{i\tau H_\text{PXP}}e^{-i\tau H_\text{PXP}}=\mathds{1}$; 
this implies  perfect subharmonic revivals across the {\it entire} Hilbert space. 
However, upon deviating from $\theta = \pi$, we find that such revivals quickly damp out for typical initial states without N\'eel order, see Fig.~\ref{Fig:pheno} and~\cite{SOM}. 

In contrast, long-lived oscillations from the N\'eel state persist over a wide range of parameters near $\theta=\pi$ and $\tau =\tau_r/2$. To quantify the stability of oscillations and subharmonic response, we compute the
subharmonic weight $f_2(\omega_d/2) \propto |S(\omega_d/2)|^2$, defined as the normalized spectral weight of $\langle \NAB(t) \rangle$ at half the driving frequency $\omega_d=2\pi/\tau$, rescaled so $f_2(\omega_d/2)=1$ for perfect subharmonic response at $\theta=\pi$ from the N\'eel states, see~\cite{SOM}. 
The plateaus in the subharmonic weight and time-averaged entanglement entropy in Fig.~\ref{Fig:pheno}(b) signal  a persistent many-body response at frequency $\omega_d/2$ over a the broad range of parameters.

\begin{figure}[t]
\begin{center}
\includegraphics[width=0.99\columnwidth]{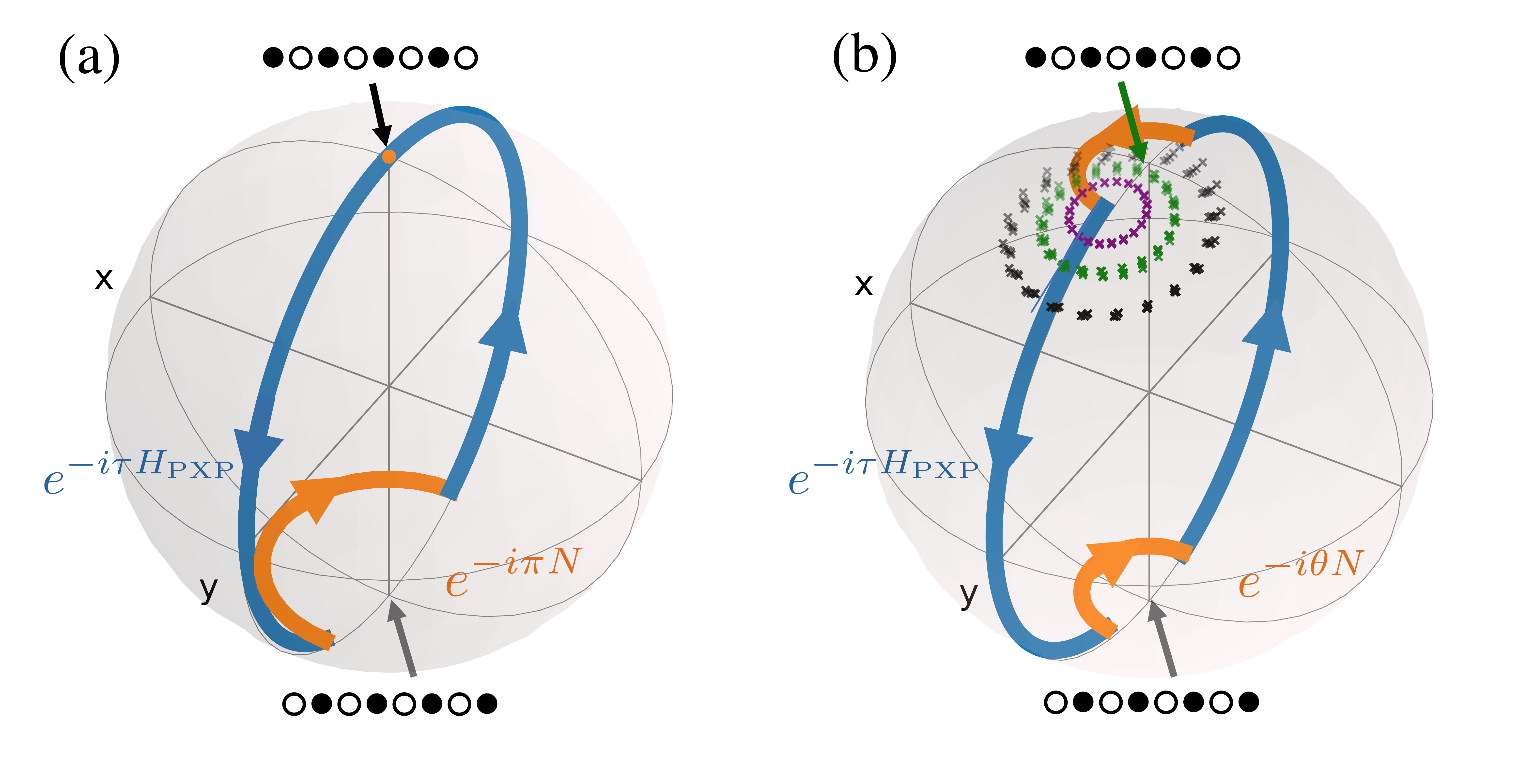}
\caption{ \label{Fig:cartoon}
Trajectories of driven PXP model for $L=16$, plotted on the Bloch sphere of the collective spin-$L/2$.
(a) The dynamics generated by two periods of $U_F(\pi,\tau)$ exhibit a perfect return to the $\ket{Z_2}$ initial state: $e^{-i\tau H_{PXP}}$ with $\tau = 0.45\,\tau_r$ under-rotates the $\ket{Z_2}$ state (blue line), then the application of $e^{-i\pi N}$ (orange line) flips the $x, y$-projections of the spin so that the second Floquet pulse completes the cycle.
(b)~The same dynamics but for $\theta=\pi - 0.05$ supports a periodic trajectory near the N\'eel state. Dynamics initialized near the periodic trajectory precess around it at stroboscopic times forming cycles depicted for 100 driving periods for three initial states~(green ring corresponds to $\ket{Z_2}$ initialization).
}
\end{center}
\end{figure}

{\it Many-body echo in su(2) subspace.}---The robustness of the subharmonic oscillation away from $\theta = \pi$ can be qualitatively understood in terms of mean-field-like trajectories on an effective Bloch sphere, by invoking the forward-scattering approximation (FSA) introduced in~\cite{Turner2017, TurnerPRB}. The FSA constructs an $L+1$ dimensional subspace that captures dynamics under $H_\text{PXP}$ from a N\'eel initial state, and approximately has the su(2) algebraic structure~\cite{Turner2017} of a spin-$L/2$ collective degree of freedom.  The $S^z$ operator is defined by the difference in excitation number on odd and even sites, $S^z = \sum_{i=1}^{L/2}(n_{2i-1}-n_{2i})$, so the N\'eel state $\ket{Z_2}$ ($\ket{Z'_2}$) corresponds to the North (South) pole. The $S^x$ operator is approximately proportional to   $H_\text{PXP}$, and generates a rotation around the $x$-axis that exchanges the two N\'eel states (blue lines in Fig.~\ref{Fig:cartoon}). Finally, $S^y$ is calculated using  su(2) commutation relations. Note that we use the  weakly deformed  PXP model~\cite{Choi2018} to generate the FSA basis, but consider dynamics under $H_\text{PXP}$~\cite{SOM}.

Computing expectation values of the collective spin operators $S^{x,y,z}$ defined above, we visualize the many-body dynamics from the N\'eel initial state under two periods of Floquet evolution~(\ref{Eq:UF}) in Fig.~\ref{Fig:cartoon}. As mentioned, $H_\text{PXP}$ implements a rotation around the $x$-axis. In contrast, the action of $e^{-i\theta N}$ pulses is more complex, since the operator $N$ does not have a closed form representation in the su(2) subspace. However, it can be approximated as $N \sim (S^z)^2$ in the vicinity of the N\'eel states $\ket{Z_2}$ and $\ket{Z_2'}$, which accumulate identical phases under $e^{-i \theta N}$, see~\cite{SOM} and Fig.~\ref{Fig:cartoon}(b).

Figure~\ref{Fig:cartoon}(a) illustrates that at $\theta = \pi$ the second application of $H_{\text{PXP}}$ returns the system to its initial state. Away from the perfect point $\theta=\pi$, trajectories from $\ket{Z_2}$ are no longer closed, but there exists a nearby closed orbit with period $2\tau$, see Fig.~\ref{Fig:cartoon}(b), explaining subharmonic response. In this picture, the existence of periodic trajectories is qualitatively similar to mean-field descriptions of time crystals~\cite{Russomanno_2017, Choi16DTC, Ho17}. However, a key difference is that the emergent spin-$L/2$ degree of freedom is not composed of independent spins, evinced by non-trivial entanglement oscillations. 
Furthermore, dynamics outside of collective spin-$L/2$ subspace are ergodic, leading to rapid thermalization from other initial states.

Even though the trajectory from $\ket{Z_2}$ is not closed, the stroboscopic dynamics (with period $2\tau$) exhibit precession around the periodic trajectory, forming islands of stability similar to Kolmogorov-Arnold-Moser tori in dynamical systems. The precession leads to characteristic beatings with an emergent timescale $T_b$,   corresponding to the period of motion on the circles around the fixed point in Fig.~\ref{Fig:cartoon}(b). The  robust subharmonic response and emergence of a beating timescale $T_b$ are a qualitative prediction of the spin-$L/2$ picture that will be confirmed below. Despite capturing much of the observed phenomenology, the spin-$L/2$ picture treats the many-body dynamics within an $L+1$ dimensional subspace. To explain how driving reduces quantum thermalization, we consider the many-body Floquet unitary.

{\it Prethermal analysis and effective Hamiltonian.}---We analyze the many-body dynamics by expanding around the perfect echo point $\theta = \pi$ where the Floquet unitary is denoted ${\cal X}_{\tau}=U_F(\pi,\tau)$. This allows us to write   $U_F(\theta,\tau) = e^{i \epsilon N}{\cal X}_{\tau}$, where $\epsilon = \pi - \theta$ is a small parameter quantifying the deviation from the perfect point. 
Since ${\cal X}_\tau^2=\mathds{1}$, this unitary is in the canonical time crystal form~\cite{Else16, vonKeyserlingk2016}, which was rigorously analyzed by~\cite{Else2017_Prethermal,ElseHo2020}, and we extend their results to the present case. 
In~\cite{SOM} we show that the Floquet unitary can be approximated by $U_F \approx {\cal V} e^{-i \epsilon H_F} {\cal X}_{\tau} {\cal V}^{\dagger}$, where ${\cal V}$ is a unitary frame transformation perturbatively close to the identity and $H_F$ is an effective Hamiltonian also constructed perturbatively in~$\epsilon$. As the DTC phenomenology depends on spectral properties of the effective Floquet unitary, which are not affected by ${\cal V}$,  we base our analysis on the leading order effective Hamiltonian and Floquet unitary, 
\begin{equation}\label{Eq:EffFloq}
H_F^{(1)} = -\frac{1}{2} \left(N + {\cal X}_\tau N {\cal X }_\tau\right) , \ U_F^{(1)}(\theta,\tau) \equiv e^{-i \epsilon H_F^{(1)}} {\cal X}_\tau.
\end{equation}
$H_F^{(1)}$ has an intuitive form, corresponding to the average Hamiltonian in a frame co-rotating with ${\cal X}$~\cite{SOM}. Thus, at leading order, $\epsilon$ sets the timescale of dynamics in the rotating frame.

A key feature of this result is that the effective Hamiltonian $H_F$ has an emergent $\mathds{Z}_2$ symmetry $[H_F,{\cal X}_\tau] = 0$ (even beyond the lowest order $H_F^{(1)}$, see~\cite{SOM}) guaranteed as long as the time-periodicity of the drive is respected. $H_F$ can rigorously be shown to accurately describe the system at least up to the prethermal timescale $T_p \gtrsim (\tau/\epsilon)e^{c_p/\epsilon}$ for some constant $c_p > 0$~\cite{Abanin_2017_CMP,Else2017_Prethermal}. 

\begin{figure}[t]
\begin{center}
\includegraphics[width=0.99\columnwidth]{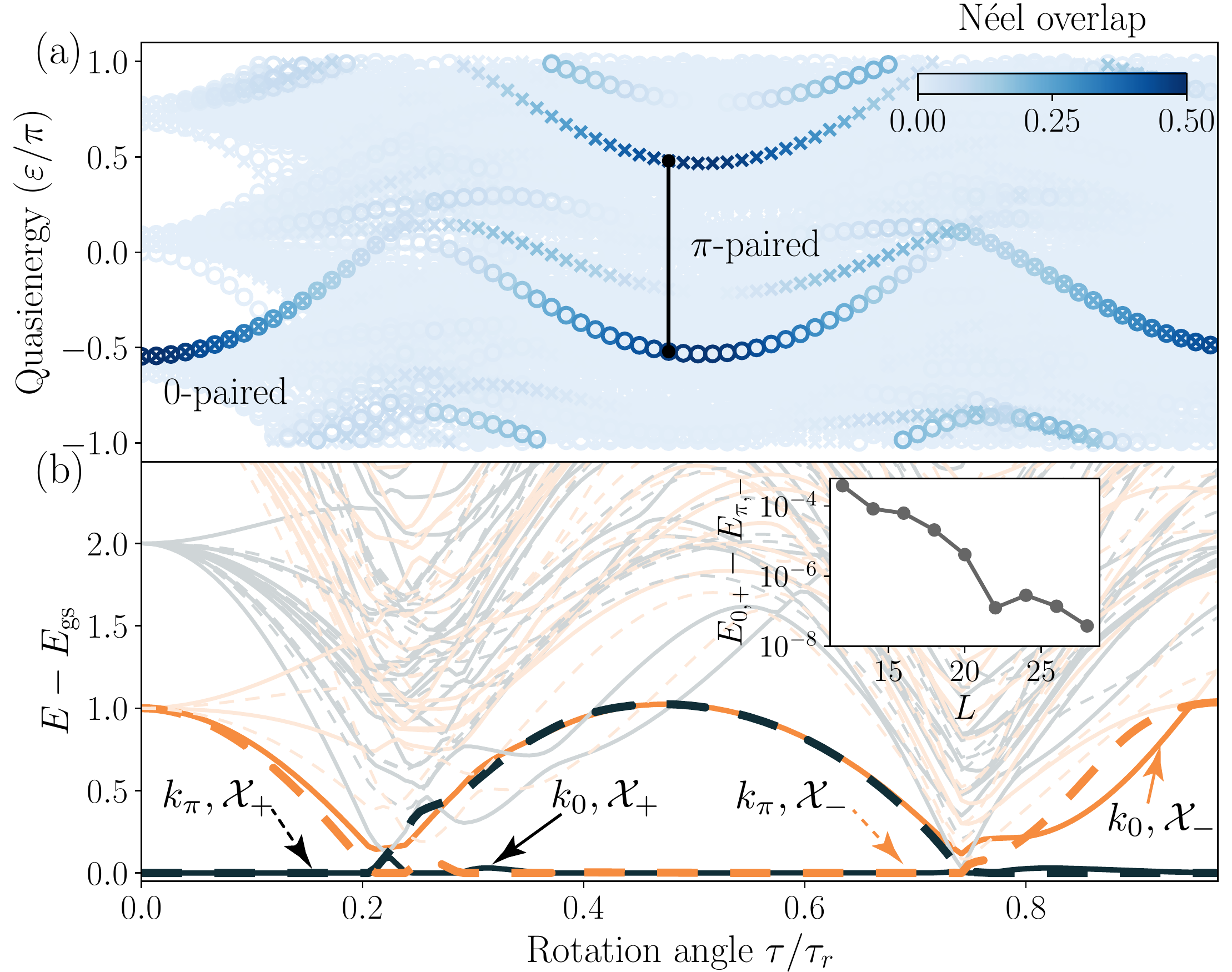}
\caption{ \label{Fig:prethermal}
(a) Eigenspectrum of $U^{(1)}_F$ plotted for various rotation angles $\tau$, $L=16$, and $\epsilon=1$, with color intensity that corresponds to overlap with N\'eel state. In the region of $\tau/\tau_r \approx 1/2$, the states with largest overlap exhibit $\pi$-pairing, indicating subharmonic response. However, near $\tau/\tau_r \approx 0,\, 1$, they exhibit $0$-pairing.
(b) Low-energy spectrum of $H^{(1)}_F$ reveals double degeneracy between ground states from $k=0$ and $\pi$ momentum sectors (denoted as $k_{0,\pi}$) with ${\cal X}_\tau$ eigenvalues $\pm 1 ({\cal X}_\pm)$ in the region that corresponds to $\pi$-pairing in (a). The splitting of the ground state manifold vanishes exponentially with system size within the $\pi$-pairing region, with the inset showing finite size scaling at $\tau/\tau_r=1/2$.
}
\end{center}
\end{figure}

The origin of the subharmonic response can be understood by analyzing eigenstates of the transformed Floquet unitary in Eq.~(\ref{Eq:EffFloq}) and their dimensionless quasi-energies $\varepsilon$, defined by $U_F^{(1)}\ket{u} = e^{i \varepsilon}\ket{u}$. 
For $\tau$ near an integer multiple of $\tau_r/2$, the Floquet operator has a pair of eigenstates characterized by strong overlap with $\ket{Z_2}, \ket{Z_2'}$, and featuring nearly degenerate quasi-energies ($\tau = 0, \tau_r)$ or quasi-energies separated by $\pi$ ($\tau = \tau_r/2)$, see Fig.~\ref{Fig:prethermal}(a). For the latter, this indicates a subharmonic response for local observables, in dynamics launched from the $\ket{Z_2}$ state. These observations imply the eigenstates can be well approximated by the long-range correlated ``cat'' states $\ket{\pm} = (\ket{Z_2} \pm \ket{Z_2'})/\sqrt{2}$ as these states carry definite momentum $k_0$ ($\ket+$) and $k_\pi$ ($\ket-$), and underlie spontaneous symmetry breaking (SSB) of the system's translation symmetry.
However, the emergent symmetries ${\cal X}_\tau$ also play a crucial role, as the $\pi$ and $0$ quasi-energy gaps occur when ${\cal X}_{\tau}$ either exchanges the two N\'eel states ($\tau=\tau_r/2$) or leaves them invariant ($\tau=0,\tau_r$). At the level of the effective Hamiltonian $H_F^{(1)}$, these $\pi$($0$)-paired eigenstates correspond to degenerate ground states in Fig.~\ref{Fig:prethermal}(b), separated by a finite gap $\Delta$ to excited states, and belonging to different (same) symmetry sectors of ${\cal X}_{\tau}$. Hence ${\cal X}_{\tau}$ symmetry breaking in the ground state is linked to DTC order and the subharmonic oscillations of spatial order~\cite{Else16,vonKeyserlingk2016, SOM}.

We argue the observed region with DTC order descends from a model with conjectured perfect scars~\cite{Choi2018, SOM}.
Specifically, if we deform the PXP model as described in ~\cite{Choi2018}, ${\cal X}_\tau$ at $\tau = \tau_r/2$ exactly exchanges the N\'eel states, and $\ket{\pm}$ become true ground states of $H_F^{(1)}$ with a constant gap $\Delta \geq 1$.
The PXP model, as well as driving for $\tau$ away from $\tau_r/2$, are weak deformations of this drive.
However, these deformations do not preserve the emergent symmetry ${\cal X}_\tau$ at the level of $H_F^{(1)}$, and could destroy the ground state degeneracy.
In Ref.~\cite{SOM}, we argue that since the emergent symmetry changes slowly as we deform the drive, the ground states throughout the $\pi$-paired region in Fig.~\ref{Fig:prethermal} can be considered as adiabatically connected to $\ket{\pm}$.
Indeed, we confirm the energy splitting in the ground state of $H_F^{(1)}$ decreases exponentially with system size, see Fig.~\ref{Fig:prethermal}(b) inset, as expected for SSB.

\begin{figure}[b]
\begin{center}
\includegraphics[width=0.99\columnwidth]{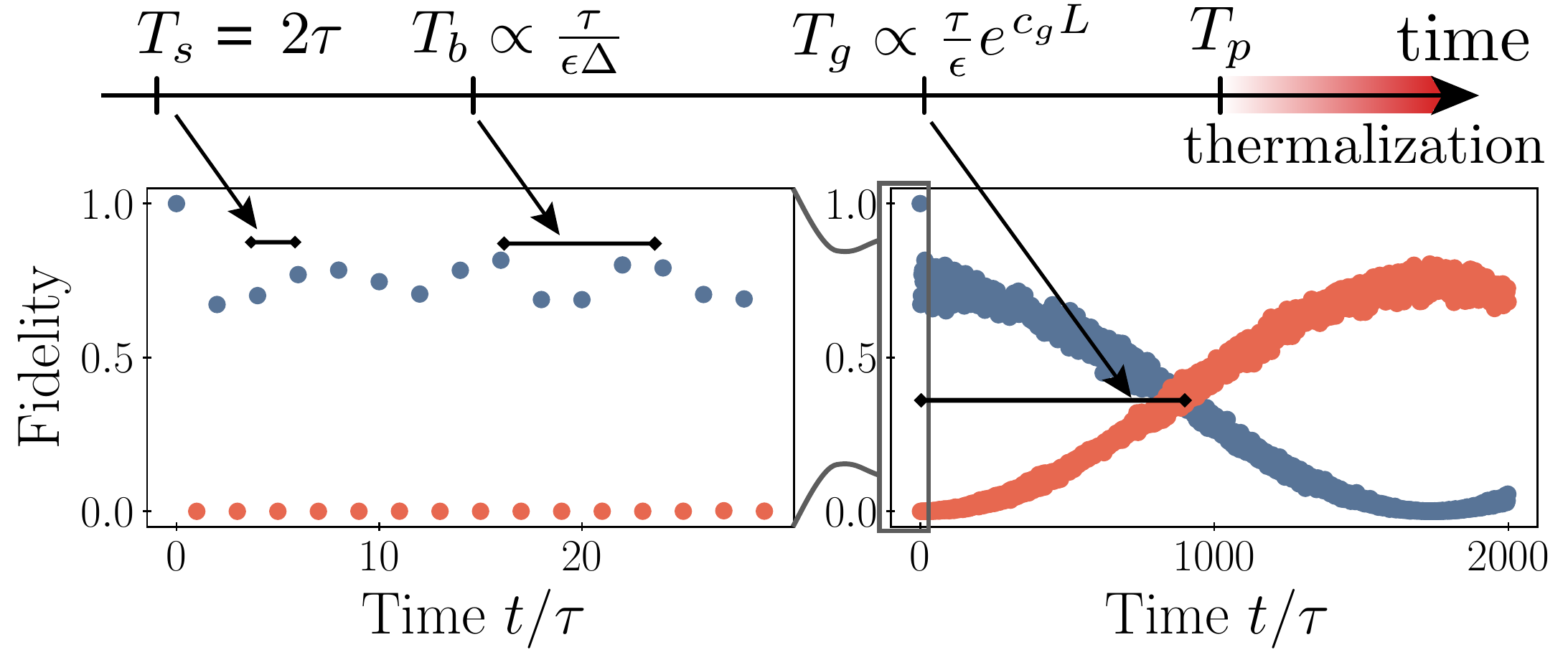}
\caption{ \label{Fig:exp} Dynamics of revival fidelity under the periodically kicked Rydberg Hamiltonian, and emergent prethermal timescales.
Stroboscopic dynamics of fidelity for $\theta=1.1\pi$, and $\tau=0.993 \tau_r/2$ reveal the subharmonic timescale $T_s$, the beating timescale $T_b$, and Rabi oscillations in the groundspace, with characteristic timescale $T_g$. 
Resonant time $\tau_r \propto 1/\Omega$ depends on the Rabi frequency.
Even (odd) multiples of $\tau_d$ are colored blue (red). Simulations performed on an $L=14$ chain.
}
\end{center}
\end{figure}

The above analysis reveals four distinct timescales emergent in the prethermal regime of Eq.~(\ref{Eq:EffFloq}). The shortest timescale $T_s = 2\tau$ is the subharmonic response. The second timescale, determined by the gap $\Delta$ in the spectrum of $H_F^{(1)}$, is $T_b\propto \tau(\epsilon \Delta)^{-1}$ and comes from overlap between the N\'eel initial state and the lowest lying excited states. Semiclassically, $T_b$ is the precession period from Fig.~\ref{Fig:cartoon}(b). Finally, the longest timescale is set by the inverse energy splitting in the ground state manifold of $H_F^{(1)}$, $T_g \propto (\tau/\epsilon)e^{c_d L}$, characteristic of SSB. All phenomenology is ultimately contingent upon the validity of the prethermal analysis, which holds until $T_p \gtrsim (\tau/\epsilon)e^{c_p / \epsilon}$. If such a bound is saturated and the system heats up to an infinite-temperature state beyond $T_p$, the physics associated with $T_g$ will become unobservable, as for fixed $\epsilon$ and large enough system sizes  $T_p < T_g$.

{\it Connections to experiments.---}We next demonstrate that the prethermal physics identified above persists beyond the idealized Floquet model~(\ref{Eq:UF}). Specifically, we replace $H_\text{PXP}$ in Eq.~(\ref{Eq:HPXP}) by the Rydberg Hamiltonian $H_\text{Ry} = (\Omega/2)\sum_i \sigma^x_i -\delta \sum_i n_i+ \sum_{i} (V_1 n_i n_{i+1}+ V_2 n_i n_{i+2})$, that includes imperfect Rydberg blockade and next-nearest-neighbor interactions. The PXP Hamiltonian is recovered from $H_\text{Ry}$ in the limit $V_1\to\infty$, $V_2=0$. Akin to the experiment in \cite{Dolev20}, we consider a 1D chain with  $V_2 = V_1/2^6$, $V_1 = 10\, \Omega$, and   choose $\delta = V_2$ to cancel the static background from the next-nearest-neighbor interactions~\cite{Dolev20}.

Figure~\ref{Fig:exp} illustrates the timescales $T_{s}$, $T_b$, and $T_g$ from stroboscopic dynamics of the revival fidelity $F_n = |\langle Z_2|U_F(\theta, \tau)^n | Z_2\rangle|^2$ generated by the kicked Hamiltonian $ H(t)=H_\text{Ry}+ \theta N \sum_k \delta({t-k\tau})$. At short times, on the order of tens of driving cycles, we observe a robust subharmonic response  at half the driving frequency, and an emergent beating timescale $T_b$. After a few hundred driving cycles, the fidelity for even periods $F_{2n}$ starts to decrease, while simultaneously for odd periods $F_{2n+1}$ starts to increase. To understand this behavior, we consider evolution at stroboscopic times and in the rotating frame, where the two nearly degenerate ground states of $H_F$, $\ket{\pm}$, form an effective two-level system with energy splitting $\Delta E = E_+ - E_-$. The inital state can be expanded as $\ket{Z_2} = (\ket{+} + \ket{-})\sqrt{2}$, and after a time $T_g = \pi/(2 \Delta E)$, it evolves into a superposition $(\ket{+} + i \ket{-})/\sqrt{2}$ equivalent to $(\ket{Z_2} - i \ket{Z_2'})/\sqrt{2}$ modulo global phase, which is a macroscopic superposition corresponding to the so-called Greenberger-Horne-Zeilinger~(GHZ) state. Dynamics in the lab frame  are related by ${\cal X}_\tau$ kicks, which exchange the N\'eel states every period. Finally, the prethermal time, when all fidelities might be expected to become exponentially small in $L$ and all local observables relax, is not visible for the system sizes or times simulated.

{\it Discussion.}---These considerations demonstrate that entanglement dynamics associated with quantum many-body scars can be stabilized and steered in the periodically kicked PXP model, resulting in an evolution strongly reminiscent of prethermal DTC order. Our construction relies on the effective many-body $\pi$-pulse realized through quantum scars, which connect the two N\'eel states via an entangled trajectory, and a driving pulse that reverses the direction of time.  Similar to prethermal time crystals, the emergent  order features a robust, long-lived subharmonic response and spatiotemporal order for a range of parameters.  However, an important difference is that these signatures are present only for eigenstates which are perturbatively close to the N\'eel initial state, and require sufficiently high fidelity state preparation to be observed~\cite{Pizzi20}. Nevertheless, we demonstrate that the signatures of DTC physics survive in an experimentally relevant model, thus providing a possible explanation for recent experimental observations in~\cite{Dolev20}. Moreover, we theoretically predict new emergent timescales that could be observed in future experiments and suggested the possibility of preparing entangled GHZ state~\cite{Omran570} in driven quench dynamics.

The phenomenon described  here drastically enhances the stability of non-ergodic dynamics thus opening a large number of exciting directions. Specifically, by extending this construction to the more complicated trajectories in the PXP model that connect highly entangled states~\cite{Michailidis2020} or to quantum scars in other models~\cite{Bull2019,Chattopadhyay,Mizuta2020,SerbynQMBS20}, control over complex entanglement dynamics could be potentially implemented. From a practical perspective, there remains a number of questions related to experiments in Rydberg arrays~\cite{Dolev20}. In particular, it is desirable to understand the dynamics in two-dimensional lattices~\cite{Michailidis2D}, including the situations where two sublattices have different numbers of nearest neighbours. In particular, in higher dimensions, there exists an  intriguing possibility of realizing a true prethermal time crystal, with a finite temperature phase transition in $H_F$. It is also desirable to build a theory for higher order subharmonic responses observed in experiments~\cite{Dolev20}, and obtain analytical understanding for continuously driven models. Finally, it is important to understand if one can implement full control over the many-body dynamics within the effective spin-$L/2$ subspace. Potentially, such controlled dynamics can be utilized for applications such as robust quantum information storage and quantum metrology. 

{\it Acknowledgments.}---
We thank Dmitry Abanin, Ehud Altman, Iris Cong, Sepehr Ebadi, Alex Keesling, Harry Levine,  Ahmed Omran, Hannes Pichler, Rhine Samajdar, Guilia Semeghini, Tout Wang, and Norman Yao for stimulating discussions.  
We acknowledge support from the Center for Ultracold Atoms, the National Science Foundation, the Vannevar Bush Faculty Fellowship, the U.S. Department of Energy, the Army Research Office MURI, and the DARPA ONISQ program~(M.L., N.M, W.W.H., D.B.);
the European Research Council (ERC) under the European Union's Horizon 2020 Research and Innovation Programme Grant Agreement No.~850899 (A.M.~and M.S.);
the Department of Energy Computational Science Graduate Fellowship under Award Number(s) DE-SC0021110 (N.M.); 
the   Moore Foundation  EPiQS initiative grant no.~GBMF4306,  the National University of Singapore (NUS) Development Grant AY2019/2020 and  the Stanford Institute for Theoretical Physics~(W.W.H.); 
the NSF Graduate Research Fellowship Program (grant DGE1745303) and The Fannie and John Hertz Foundation (D.B.); and 
the Miller Institute for Basic Research in Science~(S.C.). 

\bibliography{references}

\clearpage

\onecolumngrid

\begin{center}
	\textbf{\large Supplementary material for ``\titleofthepaper'' }\\[5pt]
\end{center}
\setcounter{equation}{0}
\setcounter{figure}{0}
\setcounter{table}{0}
\setcounter{page}{1}
\setcounter{section}{0}
\makeatletter
\renewcommand{\theequation}{S\arabic{equation}}
\renewcommand{\thefigure}{S\arabic{figure}}
\renewcommand{\thepage}{S\arabic{page}}

\twocolumngrid


In this supplementary material we provide additional data for the phenomenology of Floquet model. In addition, we describe the procedure used to visualize the dynamics on the Bloch sphere corresponding to collective spin-$L/2$ degree of freedom. Finally, we provide an analytic derivation of effective Hamiltonian and discuss the stability to generic perturbations. In the last section of this supplement we discuss the subharmonic responses in realistic Rydberg Hamiltonians. 

\section{Phenomenology of driven model}

\subsection{Quantifying the subharmonic weight}
In the main text we introduced the subharmonic weight $f_2(\omega)$ used to quantify the subharmonic response of the dynamics generated by the Floquet unitary~\eqref{Eq:UF} starting from the $\ket{Z_2}$ state, defined as
\begin{align}\label{Eq:f2}
    f_2(\omega) :=  \frac{\delta \omega\vert S(\omega)\vert^2}{\int_{1/T}^{2/t_1} d\omega' \vert S(\omega'/2)\vert^2}
\end{align}
where $t_1$ is the sampling rate, $T$ is the sampling window, $\delta \omega$ is a normalization, and $S(\omega$) is the Fourier transform of $\langle \NAB(t) \rangle$ with the mean subtracted:
\begin{align}\label{Eq:Somega}
    S(\omega) :=& \int_0^T dt\, \large( \langle \NAB(t) \rangle - \overline{\langle \NAB(t) \rangle} \large) e^{-i \omega t} \nonumber\\
    \overline{\langle \NAB(t) \rangle} :=& \frac{1}{T}\int_0^Tdt\, \langle \NAB(t) \rangle.
\end{align}
In the main text, we chose the normalization $\delta \omega$ such that the exact subharmonic response in $\NAB$ at $\theta = \pi$ and $\tau = \tau_r/2$ from the N\'eel ordered states, gives a subharmonic weight of one.

Below, we discuss the motivation behind the quantity $f_2(\omega)$. We can decompose any bounded time-varying signal  in terms of its Fourier components. We will focus on  $\langle \NAB(t)\rangle$, the signal given by the excitation imbalance in time.  Rigorously, we can look at its so-called (one-sided) power spectrum, defined as
\begin{align}
    P(\omega) := \lim_{T \to \infty} \frac{2}{T} |S(\omega)|^2,
\end{align}
which tells us how much `power' (`energy' per unit time) is contained in the frequency interval $[\omega, \omega + d\omega]$.
Note that this quantity is related to the Fourier transform of the autocorrelation function of $\langle \NAB(t)\rangle$, by the Wiener-Khinchin theorem. Also, the integral over all frequencies is the total power of the signal which is assumed finite,   can be cast by Parseval's theorem as
\begin{align}
    \int_{0}^{\infty} d\omega\, P(\omega)=
\lim_{T \to \infty} \frac{1}{T} \int_{0}^{T} dt\, |\langle \NAB (t) \rangle - \overline{\langle \NAB(t) \rangle} |^2.
\end{align}
Thus, if we had infinite-time knowledge  of the signal $\langle \NAB(t) \rangle$, we would define the fraction of its spectral weight in the frequency interval $[\omega-\delta \omega/2, \omega + \delta \omega/2]$ as
\begin{align}\label{F2def}
    F_2(\omega,\delta \omega):= \frac{\int_{\omega - \delta \omega/2}^{\omega + \delta \omega/2}d\omega' P(\omega')   }{\int_0^\infty d\omega' P(\omega')  }.
\end{align}
A value  of $F_2(\omega,\delta \omega)$ being close to unity implies that the entire power of the signal is contained in the frequency interval $[\omega - \delta \omega/2, \omega + \delta \omega/2]$.

In practice, the observation time $T$ is finite, since it is {\it required} to be smaller or comparable to the prethermal timescale otherwise thermal fluctuations dominate the response. In addition, we sample $\langle \NAB(t) \rangle$ at discrete times $t_n=nT/N$, thus the integral in the denominator of Eq.~(\ref{F2def}) is effectively truncated at some high frequency cutoff $
\omega_c = 2N/T$ set by the sampling rate, and lower cutoff $\omega_0 = 1/T$ set by the finite sampling window. Utilizing relation between $P(\omega)$ and $S(\omega)$ we write our approximation for $F_2(\omega, \delta\omega)$ as 
\begin{align}
    F_2(\omega,\delta \omega) \approx f_2(\omega) = \frac{ |S(\omega)|^2 \delta \omega     }{\int_{\omega_0}^{\omega_c}d\omega' |S(\omega')|^2 }.
\end{align}
Finally, introducing discrete Fourier transform explicitly, we write the approximate subharmonic weight as: 
\begin{align}
    f_2(\omega) = \frac{ \delta \omega \left\vert  t_1\sum_{n=0}^{N} e^{i \omega t_n} \large( \langle \NAB(t) \rangle - \overline{\langle \NAB(t) \rangle} \large) \right\vert^2}{t_1\sum_{n=0}^{N} \large( \langle \NAB(t) \rangle - \overline{\langle \NAB(t) \rangle} \large)^2}.
\end{align}

\subsection{Stability to perturbations}
In this section we demonstrate the stability against small perturbations of short time dynamics generated by the Floquet unitary $U_{F}(\pi-\epsilon,\tau)$ as defined in Eq.~(\ref{Eq:UF}) of the main text, in an infinite lattice. 
In addition, we highlight the close relation between slowdown of thermalization, as witnessed by entanglement growth, and the strength of subharmonic response.
\begin{figure*}[t]
    \centering
    \setlength{\unitlength}{\columnwidth}
\begin{picture}(0,0)
\put(0.03, 0.61){(a)}
\put(0.67, 0.61){(b)}
\put(1.29, 0.61){(c)}
\end{picture}
    \includegraphics[width=1.85\columnwidth]{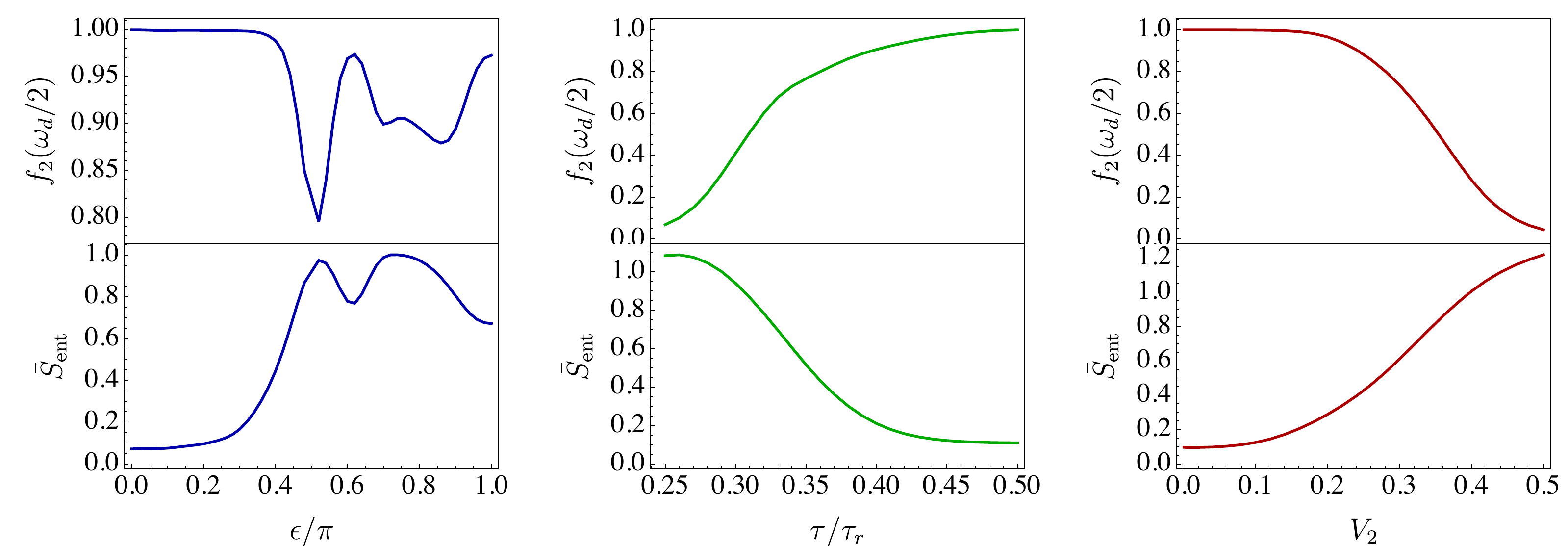}
    \caption{
    Subharmonic response for the infinite chain remains strong in a broad range of parameters (top panels) and is correlated with small entanglement production rate (bottom panels) in the dynamics under Floquet unitary~(\ref{Eq:UF}). 
    (a) The broad range of stability near the point $\epsilon=\pi-\theta =0$ originates from the ``perfect many-body echo''. The value of $\tau$ is fixed to $\tau_r/2$. 
    (b) Response of the system to driving at a frequency smaller than the natural resonance frequency $\tau_r/2$ for fixed $\theta = 0.8\pi$.
    (c) Stability with respect to perturbation of the PXP Hamiltonian by next nearest neighbor interactions for $\theta = 0.8\pi$
    and $\tau = \tau_r/2$. The total averaging time is $T=c\tau$ with $c=10$ for (a) and (c), while $c=20$ for (b).
    }
    \label{fig:perturbations}
\end{figure*}

To explore the stability of the subharmonic response we use the subharmonic weight at half the driving frequency, $f_2(\omega_d/2)$ introduced in Eqs.~(\ref{Eq:f2})-(\ref{Eq:Somega}). To quantify the rate of thermalization we compute the entanglement entropy for a bipartition of the chain $S_{\rm ent}(t)$, and calculate its time-average over $c$ driving cycles,
\begin{equation}
    \bar S_{\rm ent} = \frac{1}{c\tau}\int_{0}^{c\tau}dt\, S_{\rm ent}(t).
\end{equation}
Averages are computed over a time interval $T = c\tau$, chosen to be an integer multiple of the driving period $\tau$. 
Small values of $\bar S_{\rm ent}$ correspond to slow thermalization, whereas values of order one signal fast entanglement production. 

To this end, Fig.~\ref{fig:perturbations}(a) studies the robustness of response to increasing values of $\epsilon$. We observe that for $\epsilon/\pi < 0.3$, $\bar S_{\rm ent}$ remains constant with a small value which primarily comes from entropy generated during the micromotion. For the same $\epsilon$-interval the subharmonic response is almost maximal, indicating perfect revivals of the local Rydberg excitations.  Next, in Fig.~\ref{fig:perturbations}(b) we explore the stability when the driving period $\tau$ is changed compared to the intrinsic scar oscillation period $\tau_r/2$. A robust subharmonic response accommodated by a strong suppression of the entropy growth  is visible for $\tau  \in [0.8 ,1] \tau_{r}/2$.
The figure highlights how the subharmonic response $f_2(\omega_d/2)$ and average entropy $\bar S_{\rm ent}$ move in lockstep.

Finally,  in Figure~\ref{fig:perturbations}(c) we demonstrate the broad range of stability with respect to inclusion of next-nearest-neighbor interactions into PXP Hamiltonian, $\delta H = V_{2}\sum_i  n_i n_{i+2}$. Crucially, this perturbation does not anti-commute with the symmetry ${\cal C} = \prod_i \sigma_i^z$, ruining the perfect many-body echo point even at $\epsilon=0$. Nevertheless, similar to the previous cases, we observe that for weak enough interactions, thermalization is strongly suppressed and the subharmonic response is large. This supports our argument that the observed subharmonic response and slowdown of thermalization are closely related and both are robust against generic weak perturbations. These infinite system simulations are in agreement and further support the results for the full Rydberg Hamiltonian on finite systems in the main text, which include imperfect blockade and go out to longer times.

\subsection{Simulation Details}
Entanglement and density imbalance for infinite chains are calculated  using iTEBD simulations. The initial state is fixed to be $\ket{Z_2}$ product state. The data presented here and in the main text is obtained using a third order Trotter integrator with time step $\Delta t = \tau / 160$. At each step we truncated the smallest singular values up to a probability truncation error $\varepsilon_\text{iTEBD} \leq 10^{-6}$. 
ED simulations of both the PXP model and the full Rydberg Hamiltonian, used the Quantum Toolbox in Python (QuTiP) numerical computing package~\cite{QuTiP}. 

\section{Visualizing dynamics in SU(2) subspace}

\subsection{Construction of the spin-$L/2$ subspace}
The spin-$L/2$ subspace is constructed by repeatedly applying ladder operators $H^{\pm}$ to the N\'eel state $\ket{Z_2}$ (or $\ket{Z_2'}$). For the Hamiltonian ~(\ref{Eq:HPXP}), the ladder operators are 
 \begin{equation}
H^\pm= \sum_{i=1}^{L/2} \big[ P_{2i-1}\sigma_{2i}^{\pm}P_{2i+1}   +  P_{2i-2}\sigma_{2i-1}^{\mp}P_{2i}  \big].  
 \end{equation}
The original PXP Hamiltonian is obtained as a sum of these operators, $H_\text{PXP} = H^+ + H^-$. Repeated application of the ladder operator $H^+$ to the $\ket{Z_2}$ state would generate the $L+1$-dimensional basis of the forward scattering approximation. However, the subspace generated in such way is only an approximate spin-$L/2$ representation of su(2). Instead, we employ a weakly deformed ladder operators $\tilde{H}^{\pm}$, defined by dressing the pauli operators 
  \begin{equation}\label{Eq:Hpm}
      \tilde \sigma_i^\pm= \sigma_i^\pm \left(1+\sum_{d=2}^{n_{max}} h_d (\sigma_{i-d}^z + \sigma_{i+d}^z) \right)
  \end{equation}
where $h_d = h_0 \left(\phi^{(d-1)} - \phi^{-(d-1)} \right)^{-2}$, 
$h_0\approx 0.051$, $\phi = (1+\sqrt{5})/2$ is the golden ratio, and we use $n_{max}=8$ in our numerical simulations. The coefficients $h_d$ decay exponentially with distance and so this is a quasi-local deformation.
As demonstrated in Ref.~\cite{Choi2018}, the subspace spanned by the unnormalized vectors $\ket{Z_2}, \tilde{H}^+\ket{Z_2},\ldots (\tilde{H}^+)^{N-1}\ket{Z_2}, \ket{Z_2'}$ gives rise to a numerically \textit{exact} representation of su(2) with spin quantum number $L/2$. We used this subspace to define spin operators, but considered dynamics generated by the undeformed PXP Hamiltonian.

The operator  $S^z = \sum_{i=1}^{L/2}(n_{2i-1}-n_{2i})$ acts as the spin-$L/2$ $S^z$ operator in this space, assuming eigenvalues $-L/2,\ldots, L/2$ for a system with $L$ atoms~\cite{Turner2017,Choi2018}. The $S^x$ operator is given by $(\tilde{H}^{+} + \tilde{H}^{-})/(2\tilde\tau_r)$, where $\tilde\tau_r\approx 4.85962$ is the period of scarred oscillations under the deformed Hamiltonian. 
This is close enough to the value $\tau_r \approx 1.51\pi = 4.74280$, that we replace $\tilde{\tau_r}$ with $\tau_r$ in numerical simulations. This can be done safely since none of the observed physics depends sensitively on $\tau_r$.
Finally, the $y$-spin operator is obtained from the canonical commutation relation, $S^y = i[S^x,S^z]$. Operators $S^z$ and $S^\pm = (S^x+iS^y)/2$ defined above satisfy the canonical commutation relations, but only within the $L+1$ dimensional subspace. Specifically, let $\tilde{\cal K}$ be a projector onto the subspace generated by repeated application of the deformed ladder operators on $\ket{Z_2}$. Then, the ladder operators satisfy, up to numerical precision,
\begin{align}
    \tilde{\cal K} [S^z, S^{\pm}] \tilde{\cal K} = \pm  \tilde{\cal K} S^{\pm} \tilde{\cal K}.
\end{align}

By computing expectation values of these operators, we visualize vectors $\langle S^\alpha \rangle(t)$ as points on the Bloch sphere,  where N\'eel states $\ket{Z_2}$ ($\ket{Z'_2}$) correspond to the North (South) pole, see Fig.~\ref{Fig:cartoon}. 

\subsection{Action of $N$ in the subspace}
In the spin-$L/2$ scarred subspace, the driving term $N$ behaves as $N \sim (S^z)^2$ in the vicinity of the poles. To derive this relation, first notice that the operator $N$ behaves as an Ising interaction within the scar subspace. Lets choose the convention $n_i = (1-\sigma_i^z)/2$. Then, restricting our attention to the blockaded subspace where $n_i n_{i+1}=0$, we can rewrite $N$ as
\begin{align}\label{Sz2}
    N= \sum_{i=1}^L n_i - \sum_{i=1}^L n_i n_{i+1} = \frac{L}{2} - \frac{1}{4}\sum_{i=1}^L \sigma_i^z \sigma_{i+1}^z.
\end{align}
A similar relation can be easily derived for PXP models on any lattice. However lattices where sites have different coordination number, like in the imbalanced lattices studied in \cite{Michailidis2D,Dolev20}, have uncompensated $\sum_i \sigma_i^z$ terms.

\begin{figure}[t]
    \centering
    \includegraphics[width=0.4\textwidth]{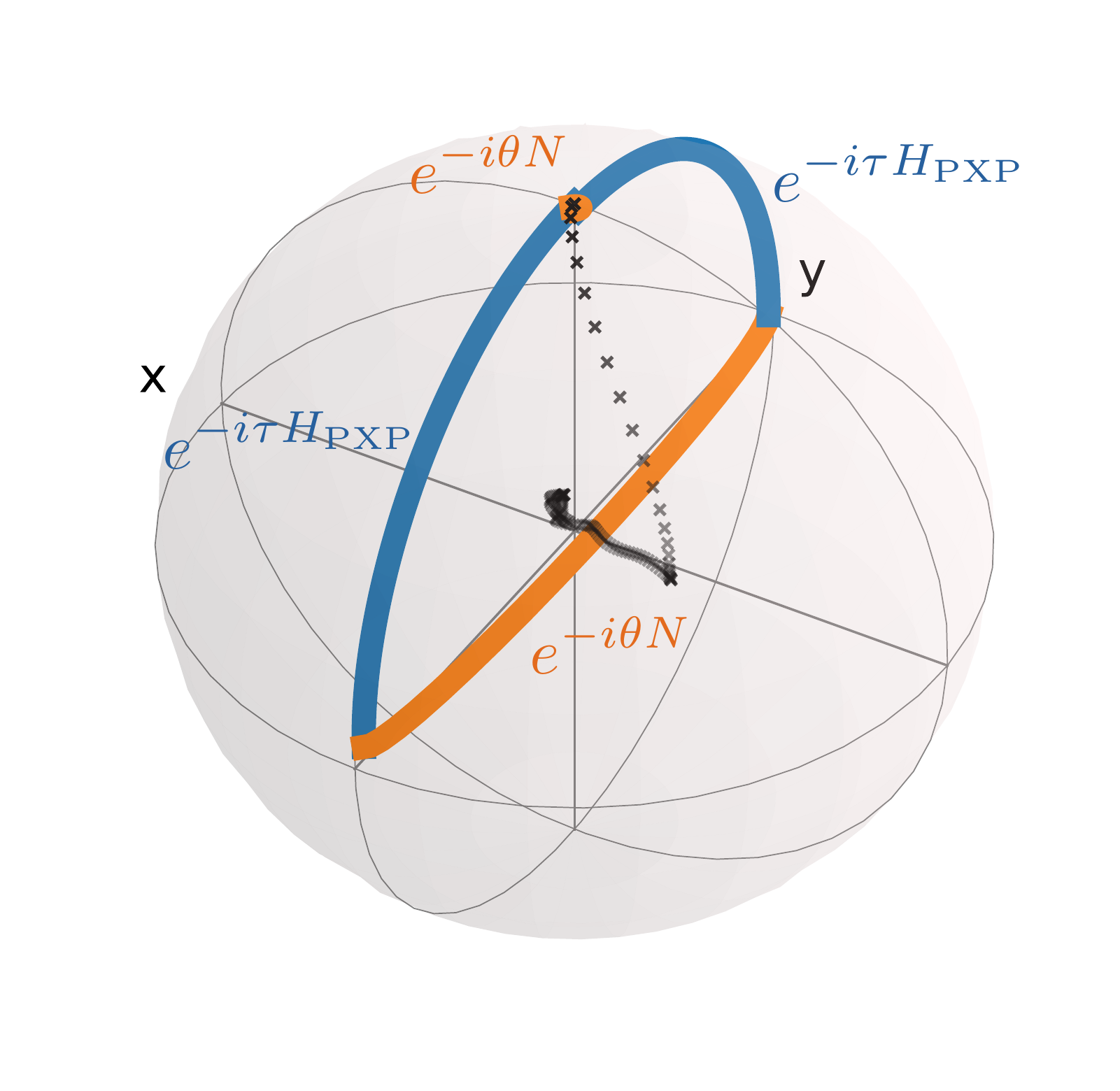}
    \caption{Pulsed driving with $\tau = \tau_r/4$, $\theta=0.985\pi$, and $L=16$ from $\ket{Z_2}$. Now $e^{-i \tau H_{\text{PXP}}}$ applies an approximate $\pi$/4 pulse, placing the state along the equator. Application of $e^{-i \theta N}$, which induces a rotation around $z$-axis near the poles, brings the state out of the collective spin-$L/2$ subspace, which is manifested as moving through the interior of the Bloch sphere. As a result, $\epsilon>0$ no longer stabilizes the trajectory, but instead induces leakage outside the manifold. At even multiples of the driving period (black crosses), the state approaches the center of the sphere, which suggests the weight of the state is leaking outside the spin-$L/2$ subspace after a few cycles and that the system is thermalizing.
    }
    \label{fig:sphere2}
\end{figure}

Next, we treat the interaction in a mean-field-like approximation. Taking into account the spatial homogeneity and N\'eel ordering, we can associate local densities with the global spin expectation value $\frac{1}{L}\langle S^z \rangle = \langle \sigma_{2i}^z \rangle = -\langle \sigma_{2i+1}^z \rangle$.
This naturally leads to the association $N \simeq \frac{1}{L} (S^z)^2$ if we take the mean-field ansatz over product states and neglect the constant in Eq.~(\ref{Sz2}). However, note that coherent states of the collective spin-$L/2$ mode~\cite{Choi2018,TurnerPRB} are not product states, in contrast to the collective mode arising from $L$ independent qubits. As a result, the mean-field argument presented above is not fully rigorous but captures crucial qualitative features.

A more quantitative way to characterize the action of the driving operator $N$ within the scar manifold is to project $N$ into the $L+1$ dimensional subspace constructed via FSA. Let $\tilde{\cal K}$ denote this projector. Then, numerically, we see that we can approximate $N$ to a high degree of accuracy by expanding in even powers of $(S^z)^2$,
\begin{align}
    \tilde{\cal K} N \tilde{\cal K} = \tilde{\cal K}\left(c_0 + c_2 \frac{1}{L} (S^z)^2 + c_4 \frac{1}{L^3} (S^z)^4 + \ldots \right) \tilde{\cal K}.
\end{align}
Odd powers of $S^z$ necessarily vanish for the periodic 1D chain by symmetry, since $N$ is invariant under translation by one site, while $S^z$ flips sign. Furthermore, off-diagonal terms like $S^x$, which change $S^z$ also necessarily vanish, since $[N,S^z]=0$.
Coefficients have been defined so that the operator $(S^z)^k/{L^{k-1}}$ is extensive, with operator norm proportional to $L$. By extrapolating to the thermodynamic limit, we see that $c_0$, $c_2$, and $c_4$ are the only non-vanishing coefficients, while higher order coefficients vanish exponentially with system size $L$.

As a result, in the spin-$L/2$ subspace, our Hamiltonian resembles a driven Lipkin-Meshkov-Glick (LMG) model which is known to support time-crystalline order. This model, studied in Ref.~\cite{Russomanno_2017}, reads $H(t) \simeq \Omega S^x + \theta \sum_{k \in \mathds{Z}} \delta({t-k\tau}) (S^z)^2$. By looking at the semiclassical limit of large $L$, Ref.~\cite{Russomanno_2017} showed that there exists stable fixed points with $2\tau$ periodic trajectories, surrounded by stable KAM tori.  However, this semi-classical limit cannot completely explain the time-crystal behavior in our model, since the mapping is only accurate if we project all of the dynamics into the $L+1$ dimensional scar subspace. Furthermore, both the driving term $N$ and the PXP Hamiltonain $H_{\text{PXP}}$ induce transitions outside this subspace, and this leakage is ignored in the LMG mapping. However, the driving term $N$ has low leakage and acts like $(S^z)^2$ in the vicinity of the poles, which is why the dynamics remain near the surface of the spin-$N/2$ Bloch sphere, for the trajectories from the N\'eel states studied here.

We illustrate in Figure~\ref{fig:sphere2} that away from the poles, $N$ does not behave like $(S^z)^2$, and instead moves states outside the spin-$L/2$ subspace. This can be seen by considering dynamics from the $\ket{Z_2}$ state when $\tau = \tau_r/4$, as the $e^{-i \tau H_{\text{PXP}}}$ pulse moves the north pole to the equator. The action of the driving pulse $e^{-i \theta N}$ is to move the state through the center of the bloch sphere. As a result, deviations from $\theta = \pi$ do not serve to stabilize this trajectory, but instead lead to thermalization, as can be seen by plotting the period $2\tau$ stroboscoipc dynamics. The timescale of this thermalization is still parametrically slow, occuring at a rate set by $\epsilon$, but this is qualitatively different from the exponentially slow thermalization observed in the gapped regime, with $\tau$ being close to an integer multiple of $\tau_r/2$. 

\section{Effective Hamiltonian and robustness to perturbations}

\subsection{Derivation of effective description}
We consider the time-periodic Hamiltonian
\begin{align}
    H(t) = H_\text{PXP} + \theta N \sum_{k \in \mathbb{Z}} \delta(t-k \tau),
    \label{eqn:H}
    \end{align}
which generates the Floquet unitary
\begin{align}
    U_F = e^{i \epsilon N} {\cal X}_\tau,
\end{align}
where ${\cal X}_\tau = e^{-i \pi N} e^{-i \tau H_\text{PXP}}$ has the property ${\cal X}_\tau^2 = 1$, and $\theta = \pi - \epsilon$.
By grouping terms in the Hamiltonian, we can rewrite $H(t)$ as
\begin{align}
    H(t) = \underbrace{H_\text{PXP} + \pi N \sum_{k \in \mathbb{Z}} \delta(t-k \tau)}_{H_0(t)} - \epsilon N \sum_{k \in \mathbb{Z}} \delta(t-k \tau). 
\end{align}
The first term $H_0(t)$ generates the propagator $U_0(t)$, which equals ${\cal X}_\tau$ after one period, and since ${\cal X}_\tau^2=1$ then $U_0(t)$ is time-periodic with double the period $U_0(t)=U_0(t+2\tau)$. Thus, we can move into a rotating frame, with respect to $U_0(t)$, and analyze the dynamics there. We assume the $-\epsilon N$ pulse is applied after the $\pi N$ pulse, so that the Hamiltonian in the rotating frame reads
\begin{multline}
    H_\text{rot}(t) = - \epsilon \Big[{\cal X}_\tau N{\cal X}_\tau \sum_{k \in \mathbb{Z}} \delta(t-(2k-1) \tau) 
    \\
    + N \sum_{k \in \mathbb{Z}} \delta(t-2k \tau)\Big],
\end{multline}
which is equivalent to the Floquet unitary $U_F(2\tau) = e^{i \epsilon N}e^{i \epsilon {\cal X}_\tau^\dagger N{\cal X}_\tau }$.
This Hamiltonian has a twisted time-translation symmetry $H_\text{rot}(t) = {\cal X}_\tau H_\text{rot}(t+\tau) {\cal X}_\tau$, immediately ensuring that time-averaged Hamiltonian is symmetric under ${\cal X}_\tau$. 

However, as discussed in the main text, we can make a stronger statement, that the ${\cal X}_\tau$ symmetry holds to all orders in a perturbative expansion, using results from Refs.~\cite{Abanin_2017_CMP, ElseHo2020}. The aim is to derive an effective description for stroboscopic dynamics. To that end we can reparameterize the Hamiltonian in the rotating frame to make the analysis easier, by introducing a new time parameter $t'$ defined over one fundamental region $t' \in [0,2\epsilon)$. The unitary time evolution operator can be written as
\begin{align}
    U(t') = U_0(t') U_1(t'),
\end{align}
where $U_0(t')$ is now $2\epsilon$ periodic and is
\begin{align}
U_0(t') = \begin{cases}
     {\cal X}_\tau \qquad  & \text{ for } 0 \leq t' < \epsilon \\
     1 \qquad & \text{ for } \epsilon \leq t' < 2\epsilon,
\end{cases}
\end{align}
and $U_1(t')$ is generated by the time-dependent Hamiltonian
\begin{align}
    H_1(t') = \begin{cases}
    -{\cal X}_\tau N{\cal X}_\tau & \qquad \text{ for } 0 \leq t' < \epsilon \\
    -N & \qquad \text{ for } \epsilon  \leq t' < 2\epsilon
    \end{cases}
\end{align}
which can be written as
\begin{multline}
    H_1(t') = -\frac{1}{2} \left({\cal X}_\tau N{\cal X}_\tau + N \right) \\
    - \frac{1}{2} \left( {\cal X}_\tau N{\cal X}_\tau - N \right) \text{sgn}\left[\sin\left( \frac{t'}{2\epsilon} \right) \right].
\end{multline}
The first term of $H_1(t')$ is the time-independent average Hamiltonian, while the second term is an oscillatory term whose time-average is 0.

Our derivation of the effective description will be based on the driven Hamiltonian $H_1(t')$,  but since the stroboscopic dynamics are equivalent, the same (stroboscopic) description will apply to the Hamiltonian we first intended to analyze, Eq.~\eqref{eqn:H}.  

The Hamiltonian $H_1(t')$ has some key properties, which enable us to derive the effective Hamiltonian $H_F$ valid up to a prethermal timescale.
First, $H_1(t')$ is a quasi-local Hamiltonian, it is a sum of bounded local terms whose amplitudes decay exponentially with size of support of the terms. The term $N = \sum_i n_i$ is clearly local, and the norm of $n_i$ is one.
For the second term, ${\cal X}_\tau N {\cal X}_\tau = \sum_i {\cal X}_\tau n_i {\cal X}_\tau$,
notice that ${\cal X}_\tau$ is generated by evolution under $H_0(t)$ for a finite time $\tau$.
Since $H_{\text{PXP}}$ has local bounded operator norm, and $N$ is an on-site operator, ${\cal X}_\tau$ only spreads operators a finite distance, with exponentially decaying support beyond this distance. This can be rigorously justified using Lieb-Robinson type bounds ~\cite{hastings2010locality}. 
Second, $H_1(t')$ still obeys a twisted time-translation symmetry
\begin{align}
    H_1(t'+\epsilon) = {\cal X}_\tau  H_1(t') {\cal X}_\tau .
\end{align}
Therefore, invoking the theorem of~\textcite{ElseHo2020}, together with prethermalization bounds from Ref.~\cite{Abanin_2017_CMP}, we can derive that there is an effective description of $U(t')$ which reads
\begin{align}
    U(t') \approx U_0(t') \mathcal{V}(t') e^{-i \epsilon H_F} \mathcal{V}(0)^\dagger,
\end{align}
where $\mathcal{V}(t')$ is a $2\epsilon$-periodic small change of frame (a unitary perturbatively close in $\epsilon$  to the identity, generated by a quasi-local Hamiltonian), and $H_F$ is an effective Hamiltoninan constructed perturbatively in $\epsilon$.

To the first couple orders in $\epsilon$ this reads 
\begin{align}\label{eq:VanVleck}
    H_F = \mathfrak{H}_0 + \frac{1}{2} \sum_{n \neq 0} \frac{\epsilon [\mathfrak{H}_n, \mathfrak{H}_{-n}]}{n \pi } + \cdots,
\end{align}
where $\mathfrak{H}_n$ are the Fourier modes of $H_1(t')$. The leading order of this expansion, $\mathfrak{H}_0$ coincides with the time-averaged Hamiltonian, and gives the expression for $H_F^{(1)}$ introduced in the main text, Eq.~(\ref{Eq:EffFloq}).

A key point to note is that $H_F$ has the symmetry $[H_F,{\cal X}_\tau] = 0$.
To see this, notice that in Fourier space, $H_1(t') = \sum_n \mathfrak{H}_n e^{i \pi n t'/\epsilon}$, the twisted time-translation symmetry reads ${\cal X}_\tau \mathfrak{H}_n {\cal X}_\tau = e^{i \pi n} \mathfrak{H}_n$,
and one can readily verify $H_F$ satisfies $\mathcal{X}_\tau H_F \mathcal{X}_\tau = H_F$ at least up to second order as given in Eq.~\eqref{eq:VanVleck} 
(but we emphasize the symmetry holds to all orders). Moreover, $\mathcal{V}(t')$ obeys also the twisted time translational symmetry, $\mathcal{V}(t'+\epsilon) = {\cal X}_\tau \mathcal{V}(t') {\cal X}_\tau$.
Therefore, our final expression for the effective description of $U_F$ reads
\begin{align}
     U_F \approx \mathcal{V}(0)  {\cal X}_\tau e^{-i \epsilon H_F}   \mathcal{V}(0)^\dagger.
    \label{eqn:effU}
\end{align}
This expression implies that stroboscopically, up to a small \textit{static} frame change, dynamics are generated by an ${\cal X}_\tau$-symmetric effective Hamiltonian $H_F$ interspersed by periodic kicks of ${\cal X}_\tau$.

\begin{figure}[t]
    \centering
    \includegraphics[width=0.49\textwidth]{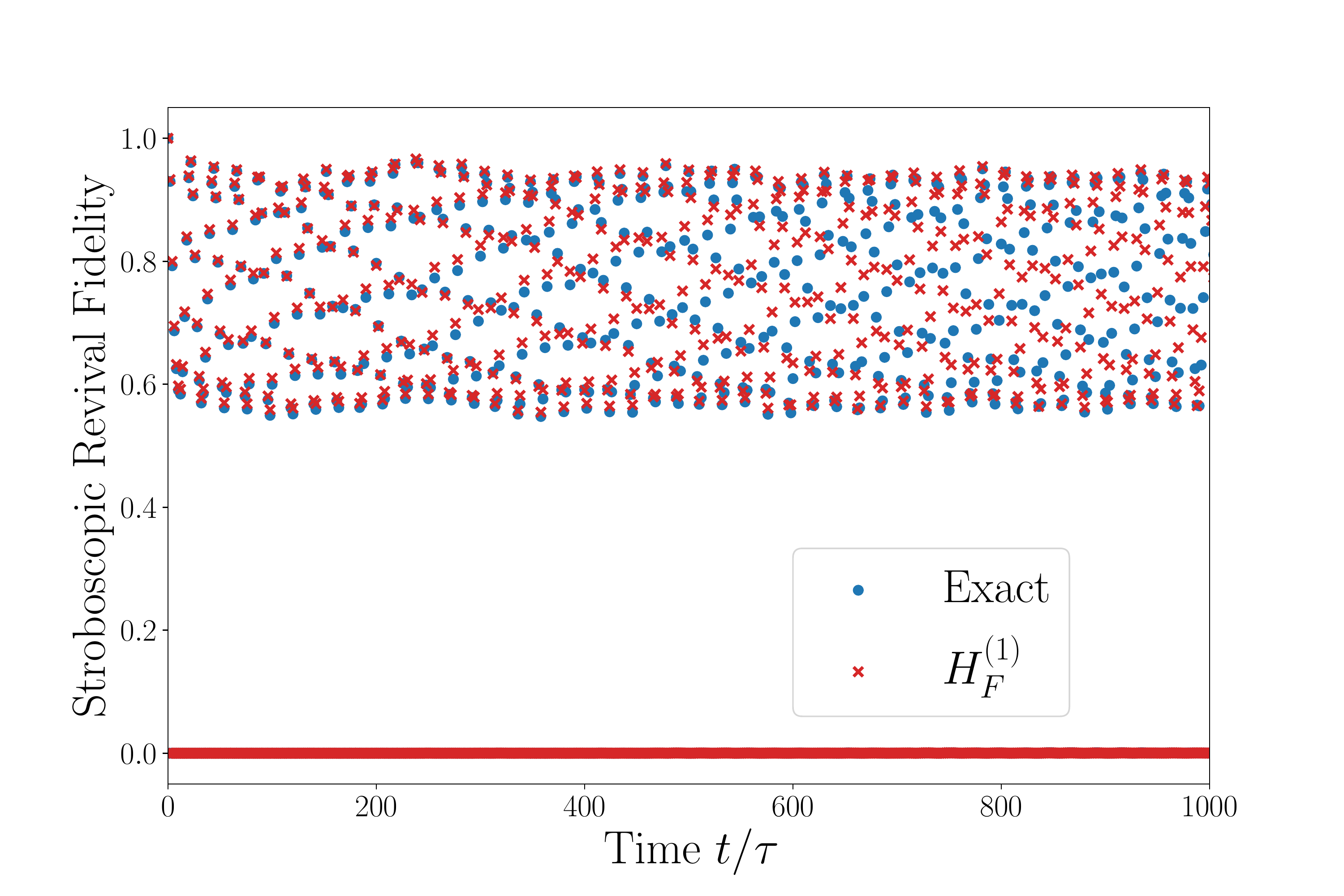}
    \caption{Stroboscopic dynamics for $L=14$, $\tau=0.55\tau_r$, $\epsilon=\pi/10$, plotted using exact floquet dynamics, and first order approximation. We see very good agreement up to long times.}
    \label{fig:UfvsHeff}
\end{figure}

This effective description lasts for at least time $t' \geq e^{c_p/\epsilon}$, in the sense that evolution under the prethermal Hamiltonian well approximates expectation values of local observables, so it describes the prethermal behavior, see Fig.~\ref{fig:UfvsHeff}. Since the original time $t$ is related to $t'$ by the factor $(\tau/\epsilon)$, it means this description lasts in original time until $t \geq (\tau/\epsilon) e^{c_p/\epsilon}$. Additionally, all time-scales generated in the time-reparameterized setting are multiplied by $(\tau/\epsilon)$ to convert back to the original time formulation. This includes the ground-state splitting oscillation time.

\subsection{Robustness to perturbations} 
Now we argue that the time-crystalline behavior is robust to arbitrary perturbations. Consider a generic time-periodic perturbation $V(t)=V(t+\tau)$ with a finite local operator norm, and a small parameter $\delta_V \ll 1/\tau$. 
The resulting  perturbed Hamiltonian is given by
\begin{align}
    H(t) &= H_0(t) + H_1(t) \\
    H_0(t) &= H_{PXP} + \pi N \sum_{k \in \mathbb{Z}} \delta(t-k \tau) \\
    H_1(t) &= -\epsilon N \sum_{k \in \mathbb{Z}} \delta(t-k \tau) + \delta_V V(t)
    \label{eqn:pert2}
\end{align}
As before, the part $H_0(t)$ generates a time-periodic frame transformation $\mathcal{X}_\tau$. However now, the local operator norm of $H_1(t)$, integrated over one period, depends not only on $\epsilon$, but also $\delta_V$. We can again write the Floquet unitary as
\begin{align}
    U(\tau) = U_0(\tau)U_1(\tau)
\end{align}
where $U_1$ is given by 
\begin{align}
    U_1(\tau) = \mathcal{T} \exp\left( -i\int_0^{\tau} dt' \underbrace{U_0(t')^\dagger H_1(t') U_0(t')}_{\tilde{H}_1(t')} \right)
\end{align}
The Hamiltonian in the rotating frame $\tilde{H}_1(t')$ is still quasi-local and has a twisted-time translation symmetry, and therefore the effective Hamiltonian will still be described by Eq.~\ref{eq:VanVleck} and have an emergent symmetry.
The only difference is that the small parameter, the integral of the local norm of $H_1(t)$ in the rotating frame, now goes as $\epsilon' = \epsilon + \delta_V \tau$. 
Hence, additional perturbations $V(t)$ reduce the prethermal timescale to $e^{c_p'/\epsilon'}$. The unitary $U_1(\tau)$ can be approximated as
\begin{align}
    U_1(\tau) \approx e^{-i \epsilon' H_1'}
\end{align}
where to leading order in $\epsilon'$ 
\begin{align}
    H_1' = \frac{-\epsilon}{2\epsilon'} (N+{\cal X}_\tau N {\cal X}_\tau) + \frac{\delta_V \tau}{2\epsilon'}(\bar{V} + {\cal X}_\tau \bar{V} {\cal X}_\tau),
    \label{eqn:H1prime}
\end{align}
and
\begin{align}
    \bar{V} = \frac{1}{\tau}\int_0^\tau U_0(t)^\dagger V(t) U_0(t) dt.
\end{align}

From Eq.~\eqref{eqn:H1prime}, we see that whether or not the ground state of this new effective Hamiltonian is connected to the unperturbed one ($\delta_V = 0$)   depends on the ratio of $\delta_V \tau/\epsilon'$. If there is a gap in the effective Hamiltonian at $\delta_V = 0$, we expect the symmetry breaking to survive as long as $\delta_V \tau \ll \epsilon$. This is why, perhaps counter-intuitively, some deviation $\epsilon$ from perfect $\pi$-pulse of $N$ is {\it required} for stability of the subharmonic response against additional arbitrary small perturbations $\delta_V V(t)$. The necessity of non-zero $\epsilon$ to protect the ground state can be seen in Fig.~\ref{fig:pulse}b, where the subharmonic response for the full Rydberg Hamiltonian disappears for a range of $\theta$ close to $\pi$, which corresponds to $\epsilon$ close to zero. 

\subsection{Origin of the many-body gap}
In this section, we present an argument for why the many-body gap observed in finite size numerics may persist in the thermodynamic limit, at $\tau = \tau_r/2$. Our analysis is specialized to 1D, as it relies on the conjecture of a deformed PXP Hamiltonian which exhibits a perfect su(2) algebra and exchange of N\'eel ordered states. 

We assume there exists a `perfect scar' Hamiltonian given by $\tilde H_\text{PXP} = \tilde H^++\tilde H^-$ from Eq.~(\ref{Eq:Hpm}). It can be viewed as a weak quasi-local deformation of original PXP Hamiltonian,
\begin{align}
    \tilde H_{\text{PXP}} = H_\text{PXP} + \delta V
\end{align}
where $\delta V = \sum_{d\geq2} h_d P_{i-1} \sigma^x_i P_{i+1} \sigma^z_{i+d}$ are long-range terms whose amplitudes decay exponentially fast, see discussion after Eq.~(\ref{Eq:Hpm}).

Using the deformed PXP model we define a corresponding $\tilde{\cal X}_{\tau}$ operator, 
\begin{align}
   \tilde{\cal X}_\tau = e^{-i \pi N} e^{-i \tau \tilde H_{\text{PXP}}},
\end{align}
which still squares to one $\tilde{\cal X}_\tau^2=1$,
and rewrite the Floquet unitary $\tilde{U}_F = e^{-i \epsilon N}\tilde{\cal X}_{\tau}$ and effective Hamiltonian $\tilde{H}_F^{(1)}=-\frac{1}{2} (N + \tilde{\cal X}_{\tau} N \tilde{\cal X}_{\tau})$. For $\tau = \tilde{\tau}_r/2$ the ``cat" states $\ket{\pm} = \large(\ket{Z_2}\pm \ket{Z_2'}\large)/\sqrt{2}$ are exact ground states of $\tilde{H}_F^{(1)}$. Furthermore, as $\tilde{\cal X}_{\tau}$ exchanges the two N\'eel states, the $\tilde{\cal X}_{\tau}$ symmetry is spontaneously broken in this ground space.

These ground states are separated from the rest of the spectrum by a constant gap $\Delta$ which is at least one. Consider first the $-N$ term. $\ket{Z_2}$ and $\ket{Z_2'}$ are the only states with $L/2$ occupied sites within the constrained subspace, and all other states have at most $L/2-1$ occupied sites, so the gap in to the lowest excitation of $-N/2$ operator is one-half. For the second term, notice that since $\tilde{\cal X}_{\tau}$ is unitary, the spectrum of $\tilde{\cal X}_{\tau} N \tilde{\cal X}_{\tau}$ is the same as $N$. Then, since $\ket{Z_2}$ and $\ket{Z_2'}$ also span the two dimensional eigenspace of $\tilde{\cal X}_{\tau} N \tilde{\cal X}_{\tau}$ with eigenvalue $L/2$, the operator norm within the subspace spanned by all remaining states is at most $L/2-1$, and we can conclude that the gap in the second term is also one-half. Therefore, the gap above the ground space in $\tilde{H}_F^{(1)}$ is lower bounded by one, i.e. $\Delta \geq 1$.

The effective Hamiltonian of the undeformed model $H^{(1)}_F$ is close to $\tilde{H}_F^{(1)}$. However the two Hamiltonians have different emergent symmetries, ${\cal X}_{\tau}$ and $\tilde{\cal X}_{\tau}$ respectively, so the `perturbation' $\tilde{H}_F^{(1)} - H^{(1)}_F$ does not respect either symmetry.
Nevertheless, using the argument in section \ref{sec:diffsyms}, we can argue the ground space of $H^{(1)}_F$ exhibits spontaneous symmetry breaking at $\tau = \tilde{\tau}_r/2$. This requires three conditions, which we justify one by one.

First, as we saw above, $\tilde{H}_F^{(1)}$ has a gapped ground state with SSB. Next, we show $H^{(1)}_F$ is a deformation of $\tilde{H}_F^{(1)}$ with finite local norm $\approx h_0 = 0.051 \ll \Delta$ that appears small enough compared than the gap, that pairing still appears in $H^{(1)}_F$. Recall $H^{(1)}_F = -\frac{1}{2}(N + {\cal X}_\tau N {\cal X}_\tau)$. In contrast, $\tilde{H}_F^{(1)} = -\frac{1}{2}(N + \tilde{\cal X}_\tau N \tilde{\cal X}_\tau)$. Then,
\begin{equation}
    \Delta H_F^{(1)} =-\frac12\Big( e^{-i \tau H_{\text{PXP}}} N e^{i \tau H_{\text{PXP}}}  - e^{-i \tau \tilde H_{\text{PXP}}} N e^{i \tau \tilde H_{\text{PXP}}} \Big)
\end{equation}
is the difference of the $N$ operator time-evolved under the two different unitaries. We expect that since these two unitaries are close to one another, $\Delta H_F^{(1)}$ should be small. This is quantified by the Duhamel expansion:
\begin{align}\label{eq:Duhamel}
    |\Delta H_F^{(1)}|  = 
    \left|\int_0^\tau ds U_1(s)^\dagger [ U_2^\dagger(\tau-s) N U_2(\tau-s), \delta V] U_1(s)\right|,
\end{align}
where $U_1(t) = e^{-i t H_\text{PXP}}$ and $U_2(t) = e^{-i t \tilde H_{\text{PXP}}}$. From here we see that, $\Delta H_F^{(1)}$ is quasilocal Hamiltonian whose local norm is set by $\delta V$, and is small as long as $h_0 \ll 1/\tau$.

The final ingredient is to show that two symmetries ${\cal X}_\tau$ and $\tilde{\cal X}_\tau$ are also close. Specifically, we show the two symmetries are related by a local unitary transformation, ${\cal X}_\tau = {\cal P}\tilde{\cal X}_\tau{\cal P}^\dagger$, where ${\cal P}$ is generated by a quasi-local Hamiltonian with small local operator norm. Recall that $\tilde{H}_{\text{PXP}}$, $H_{\text{PXP}}$, and $\delta V$ are all anti-symmetric under ${\cal C} = e^{-i \pi N}$. Thus, we can write the two symmetries as
\begin{align}
    \tilde{\cal X}_{\tau} &= {\cal C} e^{-i \tau \tilde{H}_{\text{PXP}}} = e^{i \tau \tilde{H}_{\text{PXP}}/2} {\cal C} e^{-i \tau \tilde{H}_{\text{PXP}}/2} \\
    {\cal X}_{\tau} &= {\cal C} e^{-i \tau (\tilde{H}_{\text{PXP}} - \delta V)} = e^{i \tau (\tilde{H}_{\text{PXP}} - \delta V)/2} {\cal C} e^{-i \tau (\tilde{H}_{\text{PXP}} - \delta V)/2} \nonumber
\end{align}
Now, we can treat the unitary $e^{-i \tau (\tilde{H}_{\text{PXP}} - \delta V)/2}$ by moving into the rotating frame w.r.t. $\tilde{H}_{\text{PXP}}$, specifically
\begin{align}
    &e^{-i \tau (\tilde{H}_{\text{PXP}} - \delta V)/2} = e^{-i \tau \tilde{H}_{\text{PXP}}/2} {\cal P}^\dagger \\
    {\cal P}^\dagger &={\cal T}\exp\left(\frac{i}{2} \int_0^\tau dt e^{i t \tilde{H}_{\text{PXP}}/2} \delta V e^{-i t \tilde{H}_{\text{PXP}}/2} \right) \nonumber
\end{align}
The frame transformation ${\cal P}$ is indeed generated by a quasi-local Hamiltonian with small local operator norm as long as $h_0 \ll 1/\tau$, the same dependence as $\Delta H_F^{(1)}$. Then, the argument from Sec.~\ref{sec:diffsyms} apply, and we can construct the ground state manifold of $H^{(1)}_F$ by an adiabatic deformation from $\tilde H^{(1)}_F$, with a spontaneously broken broken symmetry ${\cal X}_\tau$  at $\tau = \tilde{\tau_r}/2$. Thus, the two groundstates of $H_F^{(1)}$ should be dressed versions of $\large(\ket{Z_2}\pm\ket{{Z}_2'}\large)/\sqrt{2}$, with an energy splitting exponentially small in $L$. Note that the difference between $\tilde{\tau_r}$ and $\tau_r$ is small enough that we neglected this technical detail in the rest of the text, and simply worked at integer multiples of $\tau = \tau_r/2$.

\subsection{Stability with respect to $\tau$ deformations}
In the prior section, we explained why symmetry breaking occurs in the driven PXP model at $\tau = \tau_r/2$, even though the quantum scars are imperfect. 
In this section, we explain why the symmetry breaking persists across a range of $\tau$ near $\tau_r/2$. 
Consider the driven PXP model at two driving periods, $\tau=\tau_r/2$ and $\tau'$. If $\delta \tau = \tau - \tau'$ is small parameter, then the difference between the effective Hamiltonians and emergent $\mathds{Z}_2$ symmetries are both perturbatively small in $\delta \tau$. Then, since the ground state at $\tau$ is gapped and spontaneously breaks ${\cal X}_\tau$ symmetry, we can use the arguments from sec. \ref{sec:diffsyms} to argue why the ground state at $\tau'$ spontaneously breaks ${\cal X}_{\tau'}$ symmetry.

First, we show the two symmetries are related by a local unitary transformation. Indeed, it is simple to see that ${\cal X}_{\tau'} = e^{-i \pi N} e^{-i (\tau - (\tau - \tau')) H_{\text{PXP}}} = e^{-i \delta \tau H_{\text{PXP}}/2} {\cal X}_\tau e^{i \delta \tau H_{\text{PXP}}/2}$, where we used the fact that $H_{\text{PXP}}$ anti-commutes with ${\cal C}=e^{-i \pi N}$.

Then, we need to show the two Hamiltonians $H_F^{(1)}(\tau)$ and $H_F^{(1)}(\tau')$ are close. We see the difference
\begin{align}
    &H_F^{(1)}(\tau) - H_F^{(1)}(\tau') = \frac12 (\cal{X}_{\tau} N \cal{X}_{\tau} - \cal{X}_{\tau'} N \cal{X}_{\tau'})  \\
    &= \frac12 e^{i \tau H_{\text{PXP}}}( N  - e^{-i \delta \tau H_{\text{PXP}}}N e^{i \delta \tau H_{\text{PXP}}}) e^{-i \tau H_{\text{PXP}}}, \nonumber
\end{align}
is also perturbatively small in $\delta \tau$ using Eq.~(\ref{eq:Duhamel}).
As a result, the Floquet groundstates, corresponding to the ground states of $H_F^{(1)}$, are smoothly connected for nearby $\tau$ and remain nearly degenerate, in the regime with large spectral gap $\Delta$.

\subsection{Why   symmetry breaking survives}\label{sec:diffsyms}

It is well known that if a quantum many-body Hamiltonian $H_1$ with a symmetry ${\cal X}_1$, has gapped ground states that spontaneously break ${\cal X}_1$, then spontaneous symmetry-breaking also persists for a nearby Hamiltonian $H_2$  (that is sufficiently close in local norm), as long as the symmetry is preserved, as then the Hamiltonians' ground states can be adiabatically connected~\cite{Hastings_2005}.
In this section we present a slightly modified version of the statement: if two quantum many-body Hamiltonians $H_1, H_2$ are (i) perturbatively close, (ii)   have symmetries $\mathcal{X}_1$ and $\mathcal{X}_2$ respectively, which are also perturbatively close in a manner defined below, and (iii) $H_1$ spontaneously breaks the symmetry $\mathcal{X}_1$ in its gapped ground states, then $H_2$ exhibits spontaneously symmetry-breaking of $\mathcal{X}_2$ in its ground states; moreover, its ground states are adiabatically connected to those of $H_1$'s.
\\\\
We say that a  unitary $\mathcal{X}_1$ is perturbatively close to another unitary $\mathcal{X}_2$ if there  exists a perturbatively small, quasilocal Hermitian operator $S$  
such that
$\mathcal{X}_2 = e^{i S} \mathcal{X}_1 e^{-iS}$.
Now, we can write 
\begin{align}
    H_2 = e^{iS} H_1 e^{-iS} +  (H_2 - e^{iS} H_1 e^{-iS}).
\end{align}
This shows that $H_2$ can be written as a deformation of $e^{i S} H_1 e^{-i S}$. Crucially, by rotating $H_1$, we transformed its $\mathcal{X}_1$ symmetry into $\mathcal{X}_2$, and now all terms are symmetric in $\mathcal{X}_2$. As $e^{iS}$ is a unitary transformation which preserves spectral properties, the ground states of $e^{i S} H_1 e^{-i S}$ will spontaneously break the $\mathcal{X}_2$ symmetry by virtue of the assumption (iii) that $H_1$ spontaneously breaks the $\mathcal{X}_1$ symmetry. Now all that remains to argue for is that the deformation $H_2 - e^{iS} H_1 e^{-i S}$ is perturbatively small so that $H_2$'s ground states can be constructed adiabatically from $e^{i S} H_1 e^{-i S}$.
\\\\
To that end we note that 
\begin{align}
    H_2 - e^{i S} H_1 e^{-i S} = (H_2 - H_1) + (H_1 - e^{i S} H_1 e^{-i S} ).
\end{align}
By assumption (i) the first parenthesis on the right hand side $H_2 - H_1$ is perturbatively small; by assumption (ii) the second parenthesis is also perturbatively small [this follows from using the Duhmael expansion (S38)], and the statement follows.

\begin{figure}[t]
    \centering
    \includegraphics[width=\columnwidth]{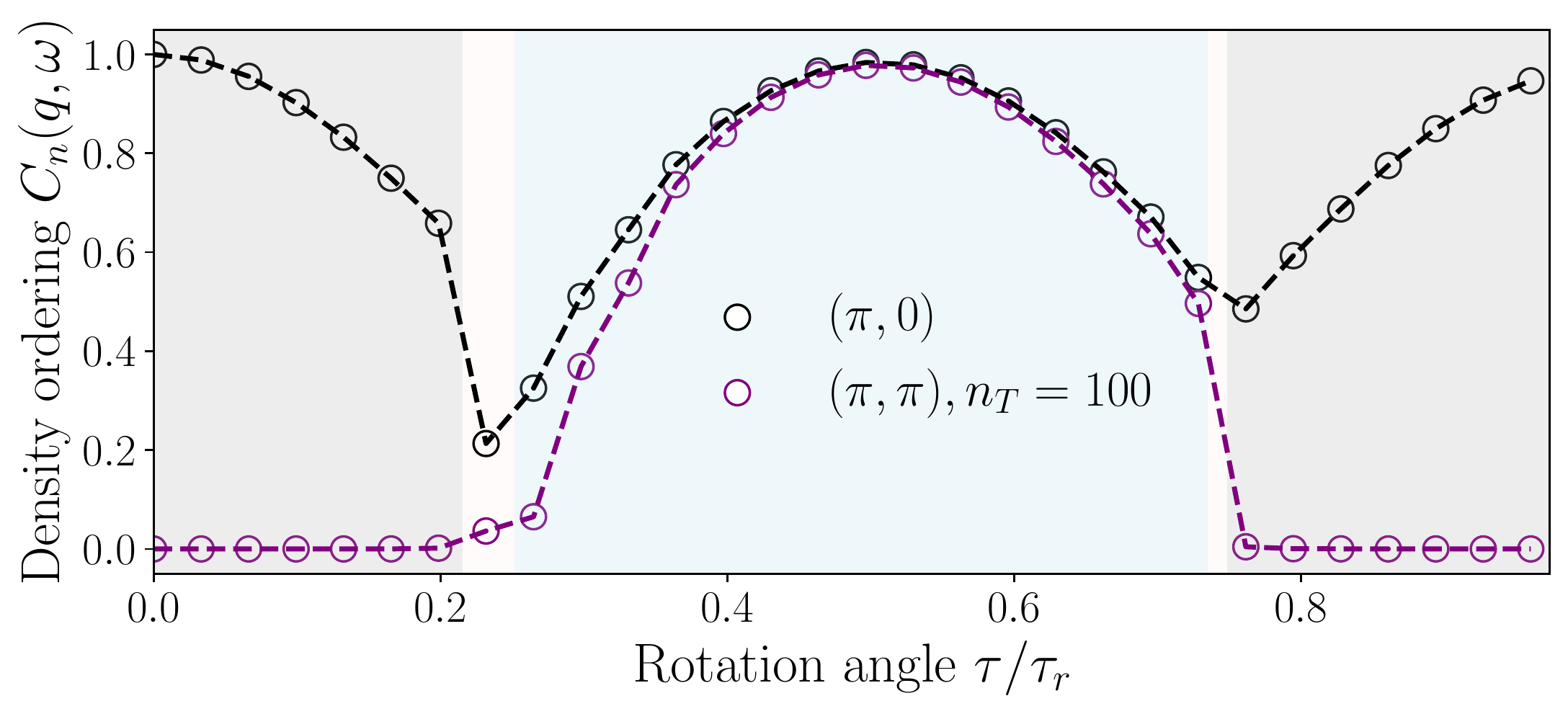}
    \caption{The Fourier transform of the density-density correlation function quantifies the order in the ground state of $H_F^{(1)}$, for $L=16$ sites. The DTC phase has non-vanishing spatiotemporal  $(\pi,\pi)$ order.}
    \label{fig:C}
\end{figure}

\begin{figure}[t]
    \centering
   \includegraphics[width=0.8\columnwidth]{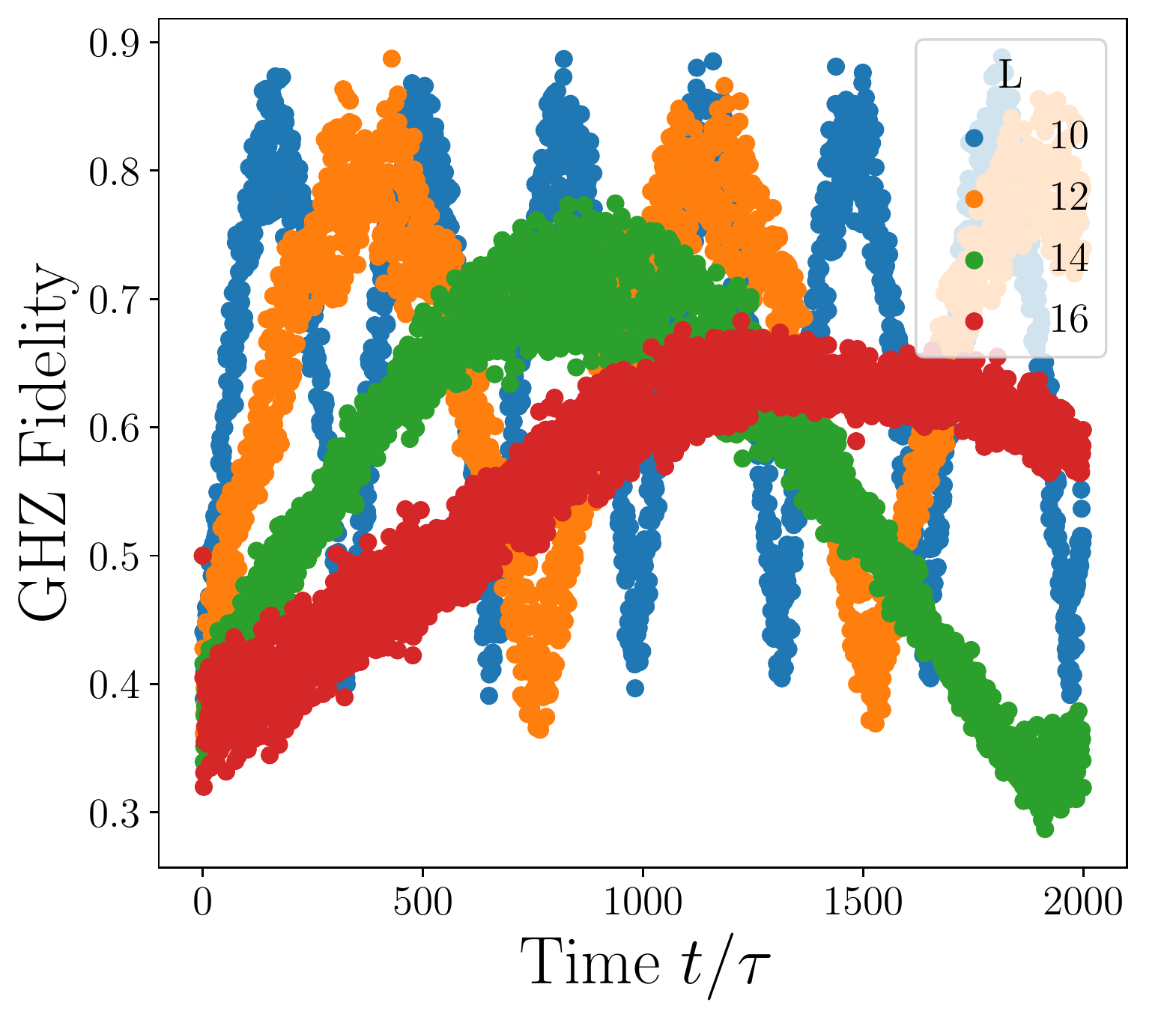}\\
   \includegraphics[width=0.8\columnwidth]{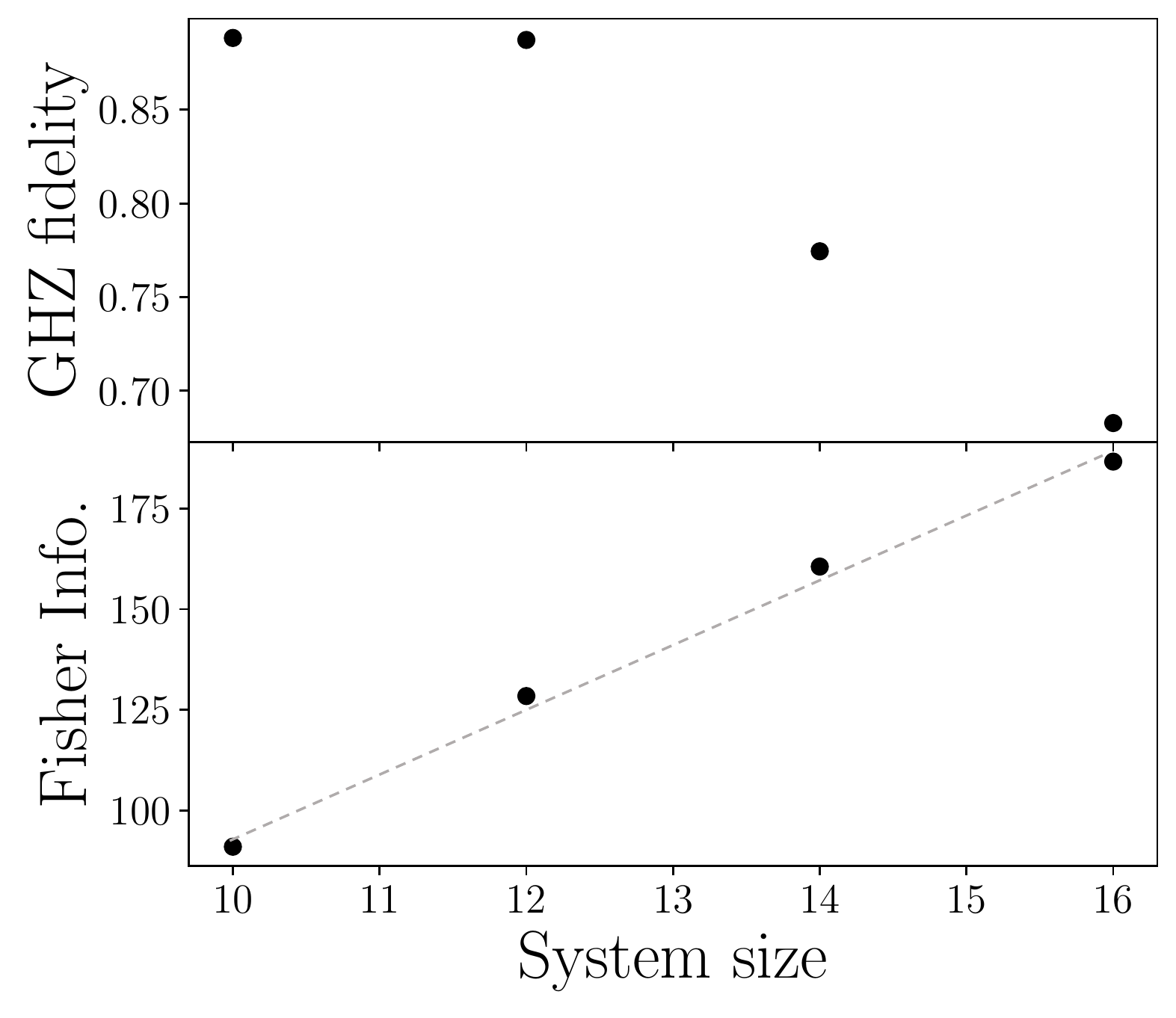}\\
    \caption{We look at pulsed driving of the Rydberg Hamiltonian, with $V_1 = 10\Omega$, $\theta = 1.1\pi$ and $\tau = \tau_r/2$. These are the same parameters of simulation as Fig.~4(a) in the main text. (top) The GHZ fidelity, as defined in Eq.~(\ref{eq:GHZ_fidelity}), reaches a maximum at time $T_g/4$, which we see is strongly dependent on system size. Furthermore, we see (middle) the fidelity of the GHZ state preparation decays with system size. As a result, the quantum Fisher information, quantifying the sensitivity of the prepared state to an applied $\NAB$ field, does not exhibit quadratic Heisenberg scaling, but instead in the regime of interest appears best described by a linear scaling, indicating the standard quantum limit. However, as we further increase system size, this scaling may continue to change.
}
    \label{fig:GHZ}
\end{figure}

\begin{figure*}[t]
\centering
   \includegraphics[width=\textwidth]{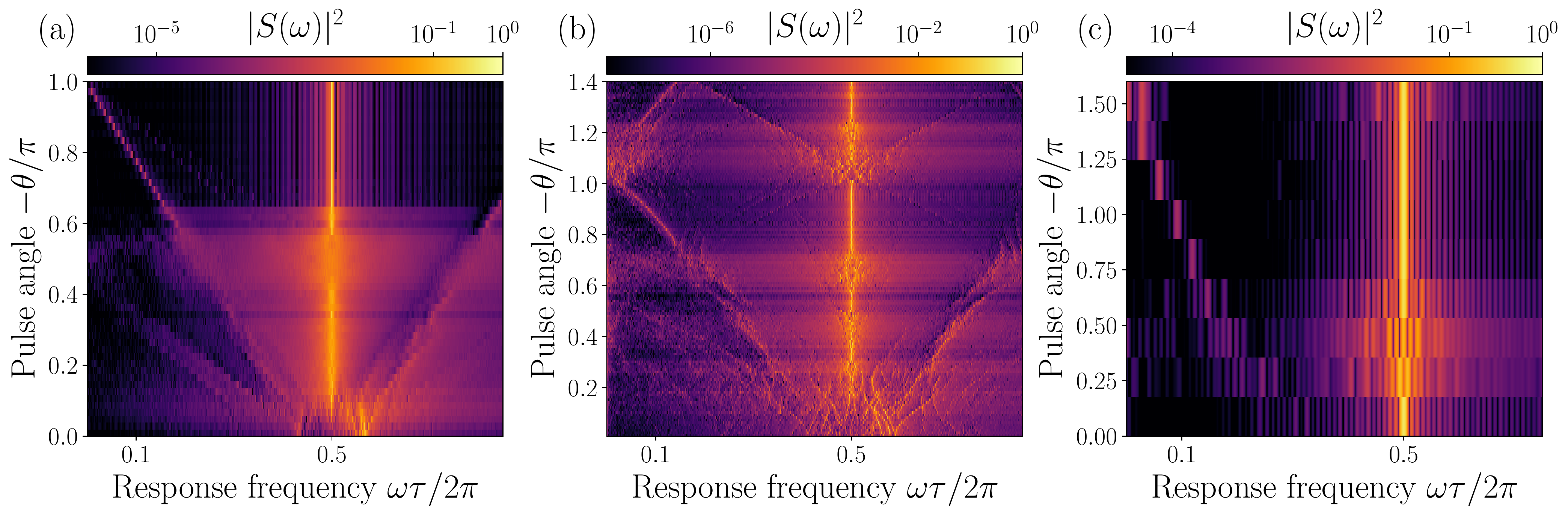}
    \caption{Fourier transforms of staggered density $\langle\NAB(t)\rangle$, for (a) the pulsed model with PXP Hamiltonian, (b) pulsed model with full Rydberg Hamiltonian, and (c) finite-width pulses with full Rydberg Hamiltonian. 
    The subharmonic locking, and emergent beating timescale are very pronounced in the ideal model. However, the signatures near $\theta=\pi$ are robust, and survive for the Rydberg Hamiltonian, even with finite-width pulses which more closely resembles the experiment~\cite{Dolev20}. 
    Notice in (b) the subharmonic response disappears in a small region around $\theta \approx \pi$. We interpret this as occurring because the gap, set by $\epsilon = \pi - \theta$, is insufficiently small to protect against perturbations coming from perturbations to the ideal model.
    In (c), notice the significant offset of the line that corresponds to the beating timescale (labeled as $1/T_b$) from $\theta = \pi$ as $\omega \rightarrow 0$. This large offset can be attributed to additional contributions to the effective Hamiltonian, introduced by finite-width pulses, see Eq.~(\ref{eq:H-fin-w}). 
    Finally, notice that for the full Ryderbg Hamiltonian, the dynamics are not equivalent under $\theta \rightarrow -\theta$.
    Data in (a) is for $\tau = 0.56\tau_r$, (b) corresponds to $\tau = 0.56\tau_r$, (c) uses $\tau = 0.5\tau_r$ with pulse width $\tau_p = 0.3\tau$. The system size is $L=24$ in (a) where we use constrained Hilbert space, whereas $L=14$ for panels (b-c) where the full $2^L$-dimensional Hilbert space is considered.
    }
    \label{fig:pulse}
\end{figure*}

\subsection{Spatiotemporal ordering in the ground state}
Finally, to characterize the ordering in the ground state $\ket{\psi_0}$ of $H_F$ via a local order parameter, we calculate a Fourier transform of the local density-density correlator
\begin{equation}
C(q,\omega) = \frac{1}{n_T L} \sum_{n=1}^{n_T}\sum_{j=1}^L e^{i (n \omega + j q)} \bra{\psi_0}  n_i U_F^{n} n_{i+j}  U_F^{-n} \ket{\psi_0},
\end{equation} 
in both spatial and temporal coordinates (restricted to stroboscopic times). 
The quantity $C(q,\omega)$ allows us to quantify different kinds of orders: the spatial (N\'eel) order, where discrete  space translation symmetry is spontaneously broken to two-site translation symmetry, is diagnosed by a large value of $C(\pi,0)$. In contrast, the spatiotemporal ordering inherent to time crystals is characterized by $C(\pi,\pi)$, which includes spontaneous breaking of discrete time translation symmetry. Figure~\ref{fig:C} illustrates that both orders are present in the range of $\tau$ where $\pi$-pairing of Floquet eigenstates persists. In contrast, in vicinity of  $\tau=0,\tau_r$ only spatial ordering is present, while the time translation symmetry remains intact.

\section{Simulation of Rydberg Hamiltonian}

\subsection{Fidelity of GHZ state preparation}
To quantify GHZ fidelity, we introduce a measure which does not depend on the phase of the GHZ state. Assume the quantum state is a pure state, and can be written as $\ket{\psi}=c_0 \ket{Z_2} + c_1 \ket{Z_2'} + ...$.
The GHZ state fidelity we compute in Fig.~\ref{fig:GHZ} is 
\begin{equation}\label{eq:GHZ_fidelity}    
\mathcal{F} = \frac12(\vert c_0 \vert^2 + \vert c_1 \vert^2 + 2 \vert c_0^* c_1\vert).
\end{equation}
When this fidelity $\mathcal{F}>0.5$, the state is verifiably entangled~\cite{Omran570}.

To understand the utility of this state preparation for metrology, we also compute the quanum Fisher information (QFI), with respect to the observable $\NAB$.
Since we focus on pure states, this is simply four times the variance of the observable, $QFI=4\langle (\Delta \NAB)^2 \rangle$.
In the ideal case, where a GHZ state is prepared, the QFI scales quadratically with the system size. This is known as Heisenberg limit scaling. 

However, the Rydberg Hamiltonian is a signficant perturbation away from the ideal pulsed model. Therefore, the groundstate of $H_F$ is only perturbatively close to $\ket{Z_2},\ket{Z_2'}$. As a result, we see the fidelity of GHZ state preparation drops noticeably with increasing system size. Indeed, we expect the fidelity to decay exponentially with system size, as typical for many-body overlap between two quantum states.
Furthermore, instead of Heisenberg limit scaling, the QFI exhibits a linear dependence, indicating standard quantum limit scaling. 

These preliminary results suggest that additional developments will be required before the time-crystalline behavior described here can be used to prepare metrologically useful states in experiments. Nevertheless, our results clearly show that for moderate system sizes, the quench dynamics reliably produce states with large overlap with GHZ states.

\subsection{Emergent timescales for finite duration of detuning pulses}
Finally, we consider an additional deformation of the pulsed driving, by applying the $N$ pulse over a finite period of time. Specifically, we use the following time-dependent Hamiltonian
\begin{equation}\label{eq:H-fin-w}
H(t) = \begin{cases}
H_\text{Ry}+ \theta N \quad \text{for} \quad  n\tau < t \leq n\tau+\tau_p/2 \\
H_\text{Ry} \quad \text{for} \quad n\tau+\tau_p/2<t\leq (n+1)\tau-\tau_p/2 \\
H_\text{Ry}+ \theta N \quad \text{for} \quad  (n+1)\tau-\tau_p/2<t\leq (n+1)\tau
\end{cases},
\end{equation}
where $n =1,2,\ldots$ is a positive integer that corresponds to the current driving period. The system evolves with Rydberg Hamiltonian at all times, though for a time $\tau_p$ it has an extra contribution from operator $N$.
This driving profile is a better approximation for the cosine driving used in Ref.~\cite{Dolev20}.
Remarkably, we see in Fig.~\ref{fig:pulse} that despite additional contributions to the effective Hamiltonian $H_F$ coming from non-commutation of $N$ and $H_{\text{Ry}}$, the subharmonic timescale $T_s$ and beating timescale $T_b$ are still clearly visible.
Furthermore, $T_b$ still exhibits linear dependence on $\theta_0 - \theta$, although the base point $\theta_0$ is no longer centered at $\theta_0=\pi$.

\end{document}